\documentclass[11pt,oneside,letterpaper]{article}
\usepackage{amssymb}
\usepackage{amsmath}
\usepackage[dvips]{graphicx}
\usepackage{setspace}
\usepackage{fancyhdr}
\usepackage{xcolor}
\usepackage{ifpdf}
\usepackage{graphicx}
\usepackage{rotating}
\usepackage{comment}
\usepackage{braket}
\usepackage{bbold}
\usepackage[pdfusetitle,bookmarks,pdfpagelabels,breaklinks,plainpages=false,pdfpagemode=UseNone]{hyperref}
\usepackage{cleveref}
 \usepackage[utf8x]{inputenc}

\usepackage{datetime,color}
\usepackage{graphicx}
\usepackage{float}
\usepackage{authblk}
\usepackage[bf,hang,footnotesize]{caption}
\usepackage[title]{appendix}

\newcommand{\be}{\begin{equation}}

\newcommand{\ee}{\end{equation}}
\newcommand{\re}{\ensuremath{{\text{Re}}}}
\newcommand{\im}{\ensuremath{{\text{Im}}}}

\newcommand{\order}[1]{{\ensuremath{\mathcal{O} \left( #1 \right)}}}
\newcommand{\eq}{{\ensuremath{{}={}}}}
\newcommand{\pheq}{{\ensuremath{\hphantom{{}={}}}}}
\newcommand{\td}{\ensuremath{{\text{d}}}}
\newcommand{\cN}{\ensuremath{{\mathcal{N}}}}

\title{\texorpdfstring{\vspace{20pt}}{}\textsc{On the Search for Multicenter AdS Black Holes from M-theory
} \texorpdfstring{\vspace{30pt}}{}}
\author[$\dag$, $\sharp$]{Ruben Monten}
\author[$\dag$, $\Diamond$]{Chiara Toldo \vspace{10mm}}

\affil[$\dag$]{\it \footnotesize  Institut de Physique Th\'eorique, Universit\'e Paris Saclay, CEA, CNRS, Orme des Merisiers, 91191 Gif-sur-Yvette Cedex, France \vspace{5mm}
}

\affil[$\sharp$]{\it \footnotesize Mani L. Bhaumik Institute for Theoretical Physics, Department of Physics \& Astronomy, University of California, Los Angeles, CA 90095, USA \vspace{5mm}}

\affil[$\Diamond$]{\it \footnotesize Institute for Theoretical Physics, University of Amsterdam, Science Park 904 Postbus 94485, 1090 GL, Amsterdam, The Netherlands}

\date{}

\begin{document}
  \maketitle
  \vspace{50pt}

    \begin{abstract}

  \vspace{5mm}
  
    \noindent
    
    We study the effective potentials for various probe branes surrounding AdS$_4$ black holes with massive halos in consistent truncations of M-theory on the Sasaki--Einstein$_7$ manifolds  \(Q^{111}\) and \(M^{111}\). These probes are either M2 branes extended in spacetime or "particle-like" probes such as internally wrapped M2 branes and, upon reduction to type IIA String theory, D6 branes corresponding to baryon operators in the dual Chern--Simons theory. We find both global and local minima of the potential outside the horizon, indicating the existence of stable and metastable multicenter AdS black holes in the extreme mass ratio regime, at fixed temperature and charges. For the planar case, we also find an instability towards nucleation of spacetime-filling M2 branes. With this analysis, we address some open questions on the holographic description of glassy phases of matter.

\end{abstract}

  \newpage
  
  \tableofcontents
      
   \section{Introduction}

Black hole bound states play an interesting role in black hole physics. A central question is how to keep a configuration of multiple black holes in equilibrium: the gravitational force should be balanced by the electromagnetic interaction and the interplay with other matter fields. Solutions of multicenter black holes in asymptotically flat spacetime were found first in the context of Einstein-Maxwell theory \cite{Majumdar:1947eu,Papapetrou:1948jw}, their causal structure and coordinate extension being studied in \cite{Hartle:1972ya}. Subsequently, asymptotically flat multicenter solutions have been found as exact solutions of the $\mathcal{N} =2$ (ungauged) supergravity equations of motion \cite{Denef:2000nb,Bates:2003vx}, and have gathered widespread attention due to their relation to  the phenomenon of wall-crossing \cite{Denef:2007vg,Kontsevich:2008fj} in the context of BPS state counting. While the first examples of multicenter black holes relied on preserving supersymmetry, it was later discovered that non-BPS multicenter black holes \cite{Goldstein:2008fq,Bena:2009ev} exist as well. Studies in the probe approximation \cite{Anninos:2011vn,Anninos:2012gk} showed that these bound states persist also at low, finite non-zero temperature.  

Finding black hole bound states in AdS spacetime is more difficult because of the confining effect of the AdS potential\footnote{The situation is different in asymptotically de Sitter spacetimes, where exact solutions of multicenter black holes in four dimensions are known, see \cite{Kastor:1992nn}. See also \cite{Sabra:2016pdt} for multicenter solutions in Euclidean signature.}. There are some indications that bound states exist: for instance hovering black holes on top of black branes in \cite{Horowitz:2014gva,Horowitz:2018coe}, and black holes in FRLW spacetimes \cite{Chimento:2013pka}. Moreover, AdS black Saturns \cite{Caldarelli:2008pz} and probe multicenter black holes \cite{Anninos:2013mfa} were constructed via approximate methods (the blackfold approach and the probe approximation, respectively). These black hole phases in AdS are particularly interesting because they model properties of strongly coupled matter via holography and allow the computation of transport coefficients that were previously beyond reach. The relation between glassy, disordered states of matter and metastable multicenter black holes was first investigated in \cite{Anninos:2013mfa}. Indeed black hole horizons  behave  like perfect fluids \cite{damour1979quelques,Price1986MembraneVO,Iqbal:2008by,Bredberg:2010ky} and in \cite{Anninos:2013mfa} it was argued that they can undergo supercooling, leading to a state that shares the disorder of a liquid and the rigidity of a solid. Several characteristics, such as qualitative behaviour of the viscosity and the relaxation dynamics of a cloud of metastable black hole probes, were checked, finding agreement with the glassy phase. Given these analogies, a proposal to model “holographic" glass by means of multicenter black holes in AdS space was put forward. 

The analysis in \cite{Anninos:2013mfa} was performed in a model of four-dimensional Fayet-Iliopoulos gauged supergravity with AdS vacuum. Embedding this setup in string theory requires taking into account additional features. For instance, in $AdS_4 \times M_6$ type IIA compactifications a linear combination of the $U(1)$s obtained by reducing the RR potentials is in fact Higgsed, see for instance the discussion in \cite{Aharony:2008ug} for the $CP^3$ reduction. The magnetic component of a Higgsed $U(1)$ generates a vortex, a magnetic flux tube confined in all but one spatial direction by the Meissner effect. At the level of string theory, this corresponds to the fact that wrapped branes carrying  D6 and  D2-charge will generically come with strings attached. Moreover, certain models coming from the reduction on $S^7$, include light charged matter with masses of the order of the AdS scale, which will condense  \cite{Gubser:2008px,Hartnoll:2008vx,Denef:2009tp}. Lastly, in analogy to their asymptotically flat counterpart, the black hole charges should correspond to wrapped D-branes. The main aim of this paper is to address these points and establish the presence of multicenter black holes in supergravity models with charged scalars, obtained by reducing M-theory on Sasaki--Einstein manifolds which have noncontractible cycles in the internal manifold, which can be wrapped by branes. Such manifolds have nontrivial Betti numbers, the $n$th Betti number parameterizing the number of linearly independent harmonic $n$-forms on the internal manifold. The charges corresponding to wrapped branes are denoted “baryonic", and “Betti multiplets" are the supermultiplets containing the (massless) gauge fields arising from the supergravity fluctuations involving the harmonic forms. The requirement that the probes carry no magnetic charge under the Higgsed $U(1)$ corresponds to the tadpole cancellation condition, which we will impose in our analysis.  

\begin{figure}[H]
\begin{center}
        \includegraphics[width=77mm]{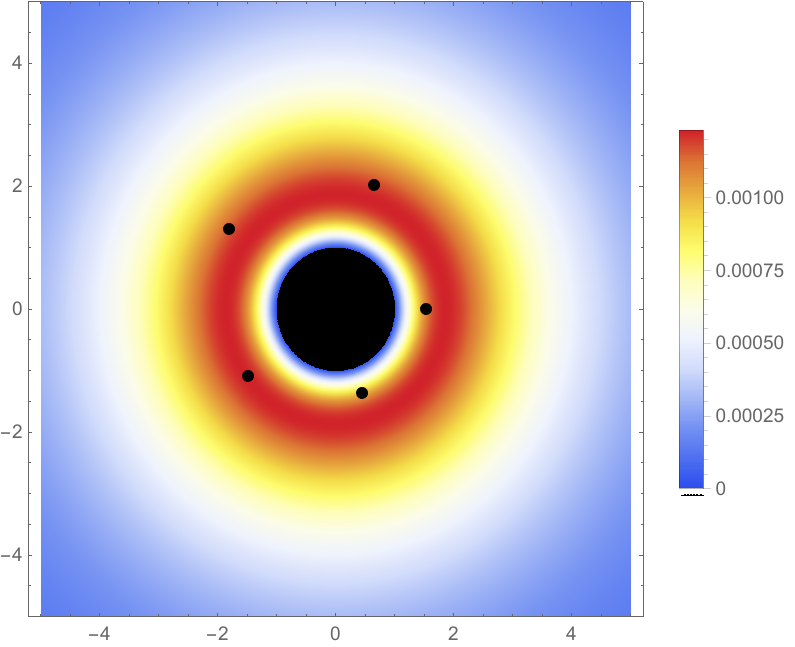}
    \end{center}
    \caption{Example of black hole with halo and probe black holes (we ignore the mutual interaction between probes) with different charges. The red region corresponds to the peak of the massive vector field: the three probes lying further away from the horizon are stable, while the other two are metastable. More details in Sec. \ref{sec_D6probes}. \label{Figure1}. }
  \end{figure}

The main result of this paper is that we find stable systems of black hole bound states in Sasaki--Einstein truncations, working in the probe approximation. We use the background black hole solutions found in \cite{Monten:2016tpu} (and generalizations thereof) which were dubbed “black holes with halos" due to the fact that their horizons are clouded by an “atmosphere" of vector field condensate. String and M- theory provides many possible objects with different charges, therefore we analyze possible instabilities of this solution towards nucleation of various kinds of M2 branes and, upon reduction to IIA, D-branes. We compute the probe brane potential using the DBI action: a global minimum of the probe potential outside the horizon signals an instability towards brane nucleation (this instability was also called Fermi seasickness in \cite{Hartnoll:2009ns}). 
Our results indicate that close to extremality, for a specific range of charges, a black hole/brane can be unstable towards brane nucleation.  Indeed, the black hole system can lower its energy by emitting probe branes, which will then sit at a minimum of the potential outside the horizon (this happens for wrapped branes) or keep moving out to the boundary of AdS (for spacetime filling branes\footnote{We use the common nomenclature "spacetime filling" M2 branes for M2 branes that are extended in $t,x,y$, hence three out of the four spacetime dimensions.}). In \cite{Klebanov:2010tj}, black brane solutions of the same model (albeit with a smaller set of charges turned on) were found to be stable towards emission of spacetime filling M2 branes and M2 branes wrapped around noncontractible cycles of the internal Sasaki--Einstein manifold. Here we consider a larger set of solutions
and show that they suffer from various nucleation instabilities. For instance, low temperature black branes with R-symmetry field turned on can emit spacetime filling branes  (see Fig. \ref{region_insta}). Of particular relevance is the fact that spherical black holes admit stable wrapped fluxed D6 probe-branes, which are particle-like objects in 4d: upon backreaction these systems can turn into multicenter black holes in AdS spacetimes. We show a black hole with halo and probes sitting outside the horizon in Fig. \ref{Figure1}. 

The paper is structured in this way: in Sec. \ref{sugra} we review the details of the 4d supergravity model obtained upon reduction of M-theory on the homogeneous 7d Sasaki--Einstein manifolds $Q^{111}$ and $M^{111}$, and we provide uplift formulae based on \cite{Cassani:2012pj}. In Sec. \ref{solutions} we describe the black hole solutions that we use as background: these include the ones found in \cite{Monten:2016tpu} for $M^{111}$ and their generalizations to the $Q^{111}$ reduction and to different horizon topologies. We discuss the numerical shooting technique used to find solutions. Sec. \ref{spacetime_filling} is devoted to the analysis of the instabilities of planar horizons towards emitting spacetime filling branes. Sec. \ref{wrapped} instead deals with the analysis of the probe stability for wrapped branes (M2 and fluxed D6). We comment on the concept of caged wall crossing and in particular on the supersymmetric  limit of the solutions. Finally, several appendices elucidate parts of our computations.

\section{Four-dimensional Supergravity setup and uplift in 11d \label{sugra}}
  
We will work with models that arise from a left-invariant consistent truncation of 11D supergravity on 7D coset manifolds of the form $M_7 = G/H$ admitting a G-invariant $SU(3)$-structure. There are a few choices of possible 7d Sasaki--Einstein manifolds that support a $\mathcal{N}=2$ supersymmetric AdS$_4$ vacuum, they are for instance listed in Table 1 of \cite{Cassani:2012pj} whose conventions we mostly follow. We choose to work with the truncations on the manifolds $Q^{111}$ and $M^{111}$, which give four-dimensional models of  Abelian $\mathcal{N}=2$ gauged supergravity. The former, $Q^{111}$, is the coset manifold $G/H$ where $G= SU(2)^3$ and $H= U(1)^2$. There are two nontrivial two-cycles, and the four-dimensional theory contains two so-called “Betti vector multiplets" in its spectrum. The truncation on $M^{111}$ used in \cite{Monten:2016tpu} can be obtained from the reduction on $Q^{111}$ upon setting to zero one of the two Betti multiplets, by suitably identifying two gauge fields $A^3=A^1$ and two scalar fields $t^3=t^1$. For the moment, we work in full generality with the $Q^{111}$ truncation. The superconformal field theory dual to the $Q^{111}$ model is the superconformal Chern--Simons flavored quiver of \cite{Jafferis:2009th,Benini:2009qs} (see also \cite{Franco:2009sp}). We proceed below by describing the content of the four-dimensional theory in the language of $\mathcal{N} =2$ supergravity.

  \subsection{4d supergravity from M-theory on $SE_7$: $M^{111}$ and $Q^{111}$ \label{4dSuGra}}
  
The field content of the four-dimensional theory consists of the gravity multiplet, the universal hypermultiplet and three vector multiplets ($n_v=3$). The conventions are spelled out in \cite{Monten:2016tpu} and we refer to that paper for details. The Lagrangian has the form
\begin{equation}
\label{N2action}
\begin{aligned}
   S =  \int \tfrac{1}{2}R\ast 1
      +{}& g_{i\bar{\jmath}}Dt^i\wedge\ast D\bar{t}^{\bar{\jmath}}
      + h_{uv}Dq^u\wedge\ast Dq^v + \tfrac{1}{4}\im\mathcal{N}_{\Lambda \Sigma}F^{\Lambda}\wedge\ast F^{\Sigma}
      + \tfrac{1}{4}\re\mathcal{N}_{\Lambda \Sigma}F^{\Lambda}\wedge F^{\Sigma}
      - V \ ,
\end{aligned}
\end{equation}
where $t^i = \tau^i + i b^i, (i = 1,2,3)$ are the three complex vectormultiplet scalars and $q^u, (u=1, \ldots 4)$ are real fields in the hypermultiplet. 
 The scalars in the vector multiplets parameterize the special K\"ahler manifold $\left(\frac{SU(1|1)}{U(1)} \right)^3$ with metric 
\begin{align} \label{GIJ}
  g_{i \bar{\jmath}} &= \partial_i \partial_{\bar{\jmath}} K (t, \bar{t}) \ , \qquad \quad \text{and}\, & K &= - \log[i( \overline{X}^{\Lambda} F_{\Lambda} - X^{\Lambda} \overline{F}_{\Lambda})] \,,
\end{align}
where $F_{\Lambda} = \partial_{\Lambda} F$ and $(X^{\Lambda}, F_{\Lambda})$ are the covariantly holomorphic sections. We will (mostly) work in the symplectic frame where all gaugings are purely electric, which is the theory obtained after dualization of the massive tensor multiplet in a massive vector multiplet (full details in \cite{Gauntlett:2009zw}). The $Q^{111}$ model is characterized by the corresponding holomorphic prepotential
\begin{equation}
  F(X) = -2i\sqrt{X^0 X^1 X^2 X^3} .
\end{equation}
The $M^{111}$ model is simply obtained by identifying two of the scalars and vectors of the $Q^{111}$ theory. Our choice of sections is such that $ X^{\Lambda} =\{X^0,X^1,X^2, X^3\}= \{1,t_2 t_3,t_1 t_3, t_1 t_2 \}$ and $F_{\Lambda} = \{i t_1 t_2 t_3, i t_1, i t_2, i t_3 \}$
and the couplings between scalars and vector fields are encoded in the period matrix  $\mathcal{N}_{\Lambda \Sigma}$, obtained via the special geometry relation
\begin{equation} \label{matrixN}
\mathcal{N}_{\Lambda \Sigma}= \overline{F}_{\Lambda \Sigma} + 2i \frac{\text{Im} F_{\Lambda \Delta} \text{Im} F_{\Sigma \Gamma} X^{\Delta} X^{\Gamma}  }{ \text{Im} F_{\Delta \Gamma} X^{\Delta} X^{\Gamma}}  \ ,
\end{equation}
with $F_{\Delta \Sigma} = \frac{\partial^2 F}{\partial X^{\Delta} \partial X^{\Sigma} }$.

The scalars in the universal hypermultiplet parameterize the quaternionic K\"ahler manifold $\frac{SU(2,1)}{SU(2) \times U(1)}$: if we denote the scalars by $q^u = (\phi,a,\xi ,\bar{\xi})$, the metric $h_{uv}$ on the manifold is
\begin{equation}
\label{univhyper}
   h_{uv} \td q^u \td q^v = \td \phi^2
       + \frac{e^{4 \phi}}{4} \left[
          \td a- \frac{i}{4}(\xi \td \bar{\xi}-\bar{\xi} \td \xi)\right]^2
       + \frac{e^{2\phi}}{4} \td \xi \td \bar{\xi} \; .
\end{equation}
The truncations under consideration are such that we gauge a U(1) isometry of the hypermultiplet manifold, and such gauging is specified by the Killing prepotentials $P_{\Lambda}^x$, from which we can read off the quaternionic Killing vectors $k_{\Lambda}^u$ \cite{Galicki:1986ja,Andrianopoli:1996cm,Andrianopoli:1996vr}
\be
\Omega_{vw}^x k_{\Lambda}^w = - \nabla_{v} P_{\Lambda}^x\,, \qquad \quad \Omega_{vw}^x = d \omega^x + \frac12 \epsilon^{xyz} \omega^y \wedge \omega^z \,.
\ee
The covariant derivatives for the vector multiplets and the hyperscalars are given by
\begin{align}
D t^i &= \td t^i + k_{\Lambda}^i A^{\Lambda} =  \td t^i  \ , & D q^u &= \td q^u + k_{\Lambda}^u A^{\Lambda}\,,
\end{align}
where we took into account that, given our gauging, only hyperscalars are charged, i.e. $k_{\Lambda}^i =0$. The explicit form of the prepotentials and Killing vectors of the gauging can be read off from \cite{Gauntlett:2009zw,Cassani:2012pj} and have the form
\begin{align}
P_0 &= 6 P_{a} -4 P_{\xi}\,, & P_1 &= 2 P_{a}\,, & P_2 &= 2 P_a\,, & P_3 &= 2 P_a \,,
\end{align}
where
\begin{align}
P_{\xi} &= \left( \begin{array}{cc}
\frac{i}{2} (1- \xi \bar{\xi} e^{-2\phi}) & -i \xi e^{-\phi} \\
-i \bar{ \xi} e^{-\phi} & -\frac{i}{2} (1- \xi \bar{\xi} e^{-2\phi}) 
\end{array}
\right) \,, 
& P_a &= \left( \begin{array}{cc}
\frac{i e^{2 \phi}}{4} & 0 \\
0 & - \frac{i e^{2 \phi}}{4}
\end{array}
\right)  \end{align}
and $P_{\Lambda} = P_{\Lambda}^x \left( -\frac{i}{2} \sigma^x \right)$. The quaternionic Killing vectors are
\begin{align}\label{killingv}
k_0 & = -6 \partial_a +4i (\xi \partial_{\xi} - \bar{\xi} \partial_{\bar{\xi}}) \,, & k_1 & = -2 \partial_a \,, & k_2 &= -2 \partial_a \, & k_3 &= -2 \partial_a.
\end{align}
With this data at our disposal, we compute the scalar potential:
\begin{equation} \label{pot_unspec}
V (t, \bar{t}, q)= (g_{i \bar{\jmath}} k_{\Lambda}^i k_{\Sigma}^{\bar{\jmath}}+4h_{u v} k^u_{\Lambda} k_{\Sigma}^v) \bar{L}^{\Lambda} L^{\Sigma}+ (f_{i}^{\Lambda} f_{\bar{\jmath}}^{\Sigma} g^{i \bar{\jmath}} - 3 \bar{L}^{\Lambda} L^{\Sigma}) P_{\Sigma}^x P_{\Lambda}^x \,,
\end{equation} 
where 
\be \label{cov_holom_sect}
(L^{\Lambda}, M_{\Lambda}) = e^{-K/2} (X^{\Lambda}, F_{\Lambda}) \equiv \mathcal{V}  \,, \qquad f_i^{\Lambda} = (\partial_i +\tfrac12 \partial_i K )L^{\Lambda}\,.
\ee

The couplings introduced by the gauging in the truncation we just described are such that one of the vectors is Higgsed. The spectrum for the truncation $Q^{111}$ can be read off from Table 7 of \cite{Cassani:2012pj} and consists of:
\begin{itemize}
\item the gravity multiplet, which contains the metric and the graviphoton;
\item a massive vector multiplet with the following content: a massive vector with $m^2 l_{AdS}^2=12$ (corresponding holographically to an operator with $\Delta =5$), which has eaten the axion $a$, and five scalars of masses $m^2 l_{AdS}^2=(18,10,10,10,4)$ corresponding to $\Delta=(6,5,5,5,4)$;
\item two Betti vector multiplets: each of them contains a massless vector and a complex scalar of mass $m^2 l_{AdS}^2 = -2> m_{BF}$, i.e. with operators with dimensions either $\Delta =(2,1)$ depending on the choice of boundary conditions.
\end{itemize}
The truncation on the manifold $M^{111}$ instead has only one Betti vector, a massive vector multiplet and the gravity one. In Sec. \ref{sec:consistentTruncation} we will see exactly which identification among the fields is required to switch off the additional Betti vector.

  \subsection{11-dimensional uplift formulae \label{uplift}}
  \label{upliftfomulae}
  
We report here the relevant details of the eleven dimensional setup from which the four-dimensional theory is obtained \cite{Cassani:2012pj}. The 11d metric reads
  \be \label{metric-ansatz}
ds_{11}^2 = e^{2V} \mathcal{K}^{-1} ds_4^2+ e^{-V} ds^2 (B_6) + e^{2 V} (\theta + A_0)^2 \,,
  \ee
   where $A_0$ is one of the four gauge fields present in the 4d Lagrangian (corresponding to the R-symmetry field), $ds_4^2$ is the line element of the 4d space (where the black hole lives), and where $\mathcal{K}$ and $V$ are
   \be \label{e2v}
   e^{2V}= \left( \sigma^2 \tau_1 \tau_2 \tau_3 \right)^{2/3}\,,
   \qquad
   \mathcal{K} = \frac18 e^{-K} = \tau_1 \tau_2 \tau_3\,,
   \ee
where we have defined $\sigma = e^{\phi}$. 

The field strength of the 11d 3-form $C_3$ is
\begin{equation*}
F_4 = d C_3 = H_4 +d B \wedge (\theta + A_0) + H_2^i \wedge \omega_i + (b^I Q_I^A \alpha_A - b^I Q_{I,A} \beta^A) \wedge (\theta +A^0) +
\end{equation*}
\be
+ Db^i \wedge \omega_i \wedge (\theta +A^0)  + D \xi^A \wedge \alpha_A -D \tilde{\xi}_A \wedge \beta^A + \chi_i \tilde{\omega}^i \,,
\label{eq:fourForm}
\ee
where $b^i$ are the axion fields, the vector $b^I$ being $b^I = (1, b^i)$, $\alpha_A$ and $\beta^A$ are 2 three-forms on $B_6$. For the solutions we will be interested in, we have that $\xi^A=0=\tilde{\xi}^A $: these are part of the hyperscalars that are set to zero. Moreover, for the specific SE$_7$ manifolds under consideration we have $Q_0^A = Q_{0,A} =0$ (see bottom of page 26 of \cite{Cassani:2012pj}). In addition, we anticipate that in all the examples used as background for the probe analysis, which are either purely electric or purely magnetic, the axions will be consistently set to zero and. The other quantities appearing in \eqref{eq:fourForm} are 
\be
\chi_i =  e_i + \mathcal{K}_{ijk} m^j b^k + Q^T_i \mathbf{C} \xi
\ee
where in our case $e_I = (e_0, e_i ) = (e_0,0,0,0)$. The four form $H_4$ is given by
\be
H_4 =  \mathcal{K}^{-1} \sigma^{4} (b^I \mathcal{E}_I + \frac12 \mathcal{K}_{ijk} m^i b^j b^k) \star_4 \mathbb{1}
\ee
which, for our purposes, will boil down to $H_4 = 6  \mathcal{K}^{-1} \sigma^{4}  \star_4 \mathbb{1}
$ and
\be \label{covdiv}
D b^i = db^i -A^0 b^j {q_{j}}^i +A^j {q_{j}}^i \,.
\ee
The quantities ${q_{j}}^i $ are related to non-Abelian gaugings and in our case are set to zero. In addition, we have
\be
H_2^i = -dA^i +2 B +b^i dA^0
\ee
and the two form B will be dualized into a scalar $a$ according to
\be 
- \sigma^{-4} \star_4 d B = da - A^I (e_I + \frac12 Q_I^T \mathbf{C} \xi) - \tilde{A}_I m^I\,,
\ee
which, in our case, will boil down to the term $- \sigma^4 \star_4 d B$ being proportional to the massive vector.

In what follows we specialize to the manifold $Q^{111}$. Here we give more specific data of the manifold. First of all, comparison with \cite{Klebanov:2010tj} where  $\psi$ is the coordinate of the fiber, tells us that
\be \label{thetadef}
\theta = d \psi + \frac14 \left( \cos \theta_1 d\phi_1 + \cos \theta_2 d\phi_2+ \cos \theta_3 d\phi_3 \right)\,,
\ee
or alternatively
 \be
d\theta = 2( \omega_1 + \omega_2 + \omega_3 )\,,
\ee
with
\be \label{omegas}
\omega_1 = \frac18 e^{12}\,, \qquad \omega_1 = \frac18 e^{34}\,, \qquad \omega_1 = \frac18 e^{56}\,,
\ee
where 
\be
e^1 = d\theta_1 \qquad e^2 = \sin \theta_1 d\phi_1
\ee
\be
e^3 = d\theta_2 \qquad e^4 = \sin \theta_2 d\phi_2
\ee
\be
e^5 = d\theta_3 \qquad e^6 = \sin \theta_3 d\phi_3
\ee
The manifold $B_6$ is specified by 
\be
ds^2 (B_6) =  \tau_1  ds_{V_1}^2+  \tau_2 ds_{V_2}^2 + \tau_3 ds_{V_3}^2\,,
\ee
 and \be
ds_{V_1}^2 = \frac18  \left( d\theta_1 + \sin^2 \theta_1 d\phi_1^2 \right) \,,  \quad ds_{V_2}^2 = \frac18  \left( d\theta_2 + \sin^2 \theta_2 d\phi_2^2 \right) \,,  \quad  ds_{V_3}^2 = \frac18 \left( d\theta_3^2 + \sin^2 \theta_3 d\phi_3^2 \right)\,.
\ee
Before wrapping up this section, let us mention that the symplectic prepotential obtained upon reduction from 11d is the so-called "STU" one, namely
\be
F_{STU}= \frac{X^1 X^2 X^3}{X^0}\,.
\ee
As done in \cite{Halmagyi:2013sla}, we will work in the rotated symplectic frame which for simplicity we call “magnetic $STU$" ($mSTU$), already introduced in Sec. \ref{4dSuGra}
\be
F_{mSTU} = -2i \sqrt{X^0 X^1 X^2 X^3} \,,
\ee
where all gaugings are electric. As explained in \cite{Gauntlett:2009zw}, this allows us to avoid working with tensor multiplets which are instead present in the $STU$  frame. Notice that the charges in the two frames are related via the symplectic rotation (see for instance \cite{DallAgata:2010ejj})
\be \label{ss_rotation}
\mathcal{S} = \left(
\begin{array}{cc|cc}
1&0&0&0\\0&0&0&- \mathbb{1}_{3 \times 3}\\ \hline
0&0&1&0\\0& \mathbb{1}_{3 \times 3} &0&0
\end{array}
\right)\ \,,  \qquad \mathcal{V}_{mSTU} = \mathcal{S} \, \mathcal{V}_{STU}\,.
 \ee
Indices $\Lambda,\Sigma= 0,1,2,3$ are used when working in the $mSTU$ frame, meanwhile $I,J= 0,1,2,3$ are used when we work in the $STU$ frame (the latter is used here and in Appendix \ref{sec:IIA}).
 
\section{Recap: \texorpdfstring{AdS$_4$}{AdS4} black holes with massive vector halo \label{solutions}}

In this section we recapitulate the main features of the static black hole solutions that will be used as the background for the probe analysis. We resort to a numerical shooting technique to find solutions of the full system of Einstein-Maxwell-scalar equations of motion. Some more details of this procedure can be found in \cite{Monten:2016tpu} where it was used to find spherical solutions of the $M^{111}$ model. We trivially extend here these technique in order to find solutions with one additional vector multiplet (arising as truncation on $Q^{111}$) and with planar and hyperbolic horizons. We explain the shooting technique below. The equations of motion can be straightforwardly computed from the Lagrangian in eq. \eqref{N2action}, and we do not report them here (those for the $M^{111}$ model can be found in Appendix A of \cite{Monten:2016tpu}).

  \subsection{Consistent truncation}
\label{sec:consistentTruncation}

In the model we have described in the previous section it is easy to see that the complex hyperscalar $\xi$ can be consistently set to zero, and the field $a$ is the Stueckelberg field which can consistently be gauged away. As explained in \cite{Monten:2016tpu}, in our quest for black hole solutions we will adopt these simplifications, therefore the quaternionic Killing prepotentials simplify considerably and their nonvanishing components are
\begin{equation} 
P_{\Lambda}^3 = (4-3 e^{2\phi}, - e^{2\phi},- e^{2\phi}, -e^{2\phi} )\,,
\end{equation}
so that the Killing vectors are 
\be
k_{\Lambda}^{a} = -(6, 2,2,2)\,.
\end{equation} 
For simplicity we have assumed a specific value for the Freund--Rubin parameter $e_0 =6$, which leads to the fixed value of AdS radius  $l_{AdS} = \tfrac12 \left(\frac{e_0}{6} \right)^{3/4}=1/2$ \cite{Halmagyi:2013sla}.

In total, we have the following matter content: three massless vector fields, a massive one, and seven scalars of masses $m^2 l_{AdS}^2 = (18,10,4,-2,-2,-2,-2)$ which correspond to dual operators of dimensions $\Delta = (6,5,4,(2,1),(2,1),(2,1),(2,1))$ where $(2,1)$ indicates the two normalizable modes for a scalar with mass $m^2 l_{AdS}^2 = -2$. 
\\
Redefining the hypermultiplet field $\phi$ as $\phi = \log \sigma$, the total action \eqref{N2action} resulting from the specified gauging is of the form
 \begin{equation} \label{eq:action}
S = \int d^4x \sqrt{-g} \left(\frac12 R - V \right) + S_V +S_H\,,
\end{equation}
where the scalar potential is, using \eqref{pot_unspec},
\be
\label{eq:potential}
V = - 8 \sigma^2 \frac{1}{\tau_1} + \frac{\sigma^4}{\tau_1 \tau_2 \tau_3} \left(\frac13 (b_1 b_2 + b_1 b_3 + b_2 b_3 + 3)^2 + (b_1 + b_2)^2 \tau_3^2 + \tau_1^2 \tau_2^2 \right) + (\text{cyclic})   \,.
\ee
For the following values of the scalar fields
\begin{equation} \label{vacua_scalar}
\tau_1=\tau_2= \tau_3 =\sigma=1\,, \qquad b_1=b_2=b_3=0\,,
\end{equation}
the potential is extremized leading to an AdS vacuum with $V_{extr} =-12$. The vector multiplet Lagrangian reads
\begin{align}
  S_V &\eq \frac14 \int{\td^4 x \sqrt{-g} \left[ - \sum_{i=1}^3 \left( \nabla(\log \tau_i) \right)^2 - \sum_{i=1}^3 \frac{(\nabla b_i)^2}{\tau_i^2}  \right]} +
  \nonumber \\
  &\pheq + \frac14 \int{\left( \im \cN_{\Lambda \Sigma} F^\Lambda \wedge *F^\Sigma + \re \cN_{\Lambda \Sigma} F^\Lambda \wedge F^\Sigma \right)} \ ,
\end{align}
with $\cN_{\Lambda \Sigma}$ given in \eqref{matrixN} (its explicit form for the models of interest is spelled out in App. A of \cite{Monten:2016tpu}). Finally, the action for the hypermultiplet sector is
\begin{equation}
  S_H = -\frac{1}{2}\int{\td^4 x \sqrt{-g} \left[2 \big(\nabla \log \sigma \big)^2 + \frac12 \sigma^4\big(\nabla a - (6 A^0 +2 A^1+2 A^2+2 A^3)\big)^2 \right]} \ ,
\end{equation}
where the scalar field $a$ is the Stueckelberg field responsible for the Higgsing of the linear combination of vector fields $6 A^0 +2 A^1+2 A^2+ 2 A^3$. Lastly, let us express the vectors  $A^{\Lambda}$ as linear combination of the massless eigenstates $\mathcal{A}_1$,  $\mathcal{A}_2$, $\mathcal{A}_3$ and the massive one $\mathcal{B}$:
\begin{eqnarray} \label{eigen_vec}
A^0 & =& \frac{1}{2}\mathcal{A}_1 +\frac{\sqrt3}{2} \mathcal{B}\,, \nonumber \\
A^1 & =& -\frac{1}{2} \mathcal{A}_1 + \frac{\sqrt3}{6} \mathcal{B}- \frac{1}{\sqrt6} \mathcal{A}_2 - \mathcal{A}_3 \,, \nonumber \\
A^2 & =& -\frac{1}{2} \mathcal{A}_1 + \frac{\sqrt3}{6} \mathcal{B}+ \frac{2}{\sqrt6} \mathcal{A}_2\, \nonumber \\
A^3 & =& -\frac{1}{2} \mathcal{A}_1 + \frac{\sqrt3}{6} \mathcal{B}- \frac{1}{\sqrt6} \mathcal{A}_2+ \mathcal{A}_3\,.
\end{eqnarray}
This action reduces to the one used in \cite{Monten:2016tpu} if we identify two of the scalars and gauge fields, $t_3 = t_1$ and $A^3 = A^1$: this corresponds to switching off the Betti vector multiplet $\mathcal{A}_3$. If instead we set $t_1 = t_2 = t_3$ and $A^1=A^2 =A^3$ we recover the action spelled out in \cite{Gauntlett:2009zw}, where the two Betti vectors $\mathcal{A}_2$ and $\mathcal{A}_3$ are set to zero. In this way we obtain the universal $SE^7$ reduction of \cite{Gauntlett:2009zw}, which coincides with the truncation on  $S^7 = SU(4)/SU(3)$ that retains the $SU(4)$ left-invariant modes.

\subsection{Ansatz for static black holes}\label{static_BH_ansatz}

We will construct numerical solutions that correspond to static black holes of the form 
\begin{equation} \label{metricR}
ds^2 = -e^{-\beta(r)} h(r) \, dt^2 + \frac{dr^2}{h(r)} +r^2 \, d\Omega_2^2 \,, \qquad d\Omega_2^2 = \left\{ \begin{array} {c}
d\theta^2 +  \sin^2 \theta d\phi^2 \qquad \kappa=1 \\
 dx^2 +  dy^2 \quad \quad \qquad \kappa=0 \\
 d\theta^2 + \sinh^2 \theta d\phi^2  \qquad \kappa= - 1\end{array} \right.
\end{equation}
which allows for asymptotically locally AdS spacetimes. We consider horizons with spherical ($\kappa =1$), planar ($\kappa =0$), and hyperbolic ($\kappa =-1$) horizons. In the latter two cases, we can obtain compact horizons by appropriately taking a quotient (e.g. Riemann surfaces of higher genus can be obtained by taking a quotient of $\mathbb{H}^2$ by a suitable subgroup).

The seven real scalar fields have only radial dependence:
\be \label{scalarR}
 \sigma=\sigma(r)\,, \quad \tau_i = \tau_i(r) \,, \quad b_i=b_i(r)\,,  \quad  i=1,2,3 \,.
\ee
The vectors are parameterized in this way: 
\begin{equation} \label{ans_vec}
\mathcal{A}_i = \xi_i(r) dt + \mathcal{P}^i d\Omega ,
\qquad \mathcal{B}  = \zeta (r) dt + \mathcal{P}^m d\Omega
\end{equation}
where $d\Omega = \cos \theta d \phi$ for $\kappa = 1$, $d\Omega = \cosh \theta d \phi$ for $\kappa = -1$, and $d\Omega = \frac12 (x dy - y dx)$ for $\kappa = 0$. As anticipated in the previous section, $\mathcal{A}_i$ are associated to actual conserved electromagnetic charges, and $\mathcal{B}$ is instead the massive vector.

The electromagnetic charges are defined as the integral of the field strength flux $F_{\mu \nu}$ and its dual $G_{\mu\nu}$ through the $S^2$ at spatial infinity:
\begin{equation}\label{chargesflux}
\mathcal{Q}_{i} = \frac{1}{4 \pi} \int_{S^2_\infty} G_{\mathcal{A}_i} \,, \qquad G_{\mu \nu, \Lambda} =  2 \frac{\partial \mathcal{L}}{\partial F^{\mu \nu, \Lambda}} \,,\end{equation}
where we have defined the dual field strengths, and
\begin{equation} \label{Fdual}
 \mathcal{P}^{i} =  \frac{1}{4 \pi} \int_{S^2_{\infty}}  F_{\mathcal{A}_i}\,.
\end{equation}
In the case of non-compact horizons, one can analogously define the charge densities as the components of the field strength and its dual along the spatial section of the conformal boundary.

The equations of motion for the vectors $\xi_i$ can be integrated by introducing the conserved quantities $p\xi_i$ which equal the asymptotic value of the charges $\mathcal{Q}_i$. The Maxwell equation imposes to the condition $ P^{\Lambda} k_{\Lambda}^u =0$, which translates into the fact that the massive vector in \eqref{ans_vec} has zero magnetic component:
\be
\mathcal{P}^m=0\,.
\ee
This is due to our ansatz \eqref{metricR}-\eqref{scalarR}-\eqref{ans_vec} which has spherical symmetry. Relaxing the condition of spherical symmetry would allow for a nontrivial magnetic component, resulting in vortex lines of the  Nielsen-Olsen type \cite{Nielsen:1973cs}, which correspond to strings ending on the D-branes which compose the background solution. 

There are other conditions we need to satisfy when the horizon is compact. First of all, the Dirac quantization conditions need to hold, since  in our symplectic frame the fermions are electrically charged:
\begin{equation} \label{dirac}
P^{\Lambda} k_{\Lambda}^u(\bar{q}) \in \mathbb{Z}\,,
\qquad
P^{\Lambda} P_{\Lambda}^3(\bar{q})  \in \mathbb{Z}\,,
\end{equation}
where $P_{\Lambda}^3 (\bar{q}) = \{1, -1 ,-1, -1 \}$ and $k_{\Lambda}^u(\bar{q})=-\{6, 2,2 , 2 \}$ are respectively the Quaternionic  Killing prepotentials and Killing vectors computed on the vacuum  \eqref{vacua_scalar}. The first Dirac quantization condition in \eqref{dirac} is automatically satisfied since we set the magnetic component of the massive field to zero, while the second, via \eqref{eigen_vec}, translates into
\begin{equation}
\label{eq:P1Quant}
2 P^1 \in \mathbb{Z} \,.
\end{equation}
Following the analysis of \cite{Gauntlett:2009dn,Bobev:2011rv}, we notice that the equations of motion and the background fields have the following scaling symmetry, 
\be
t \rightarrow \gamma \, t  \ , \qquad \beta \rightarrow \beta + 2 \log \gamma \ ,  \qquad \zeta \rightarrow \frac{\zeta}{\gamma} \,, \qquad \xi_i \rightarrow \frac{\xi_i}{\gamma}  \ ,
 \ee
which can be used to choose  $ \lim_{r \rightarrow \infty} \beta  =0 $ without loss of generality.

The equations of motion for the $M^{111}$ case are reported in full detail in \cite{Monten:2016tpu}, Appendix B, and for brevity we do not report the full $Q^{111}$ here. From the analysis of the equations of motion one can see that, in total, there are 18 degrees of freedom: two from the warp factors $\beta$ and $h$, whose equations of motion are first order. The scalars $\tau_i$, $b_i$ and $\sigma$, and the massive vector mode $\zeta$ have second order equations of motion, hence they bring additional 16 degrees of freedom as summarized in \Cref{tab:dofOverview}.

The radial coordinate of the black hole solutions is denoted by $u$, whose relation to the radial Schwarzschild coordinate is\footnote{\label{fn:planarSymmetry}For planar horizons there is another scaling symmetry which allows to pick $r_H$ =1 without loss of generality, see for instance \cite{Gauntlett:2009dn,Bobev:2011rv}.}
\begin{equation}
u= \log \left(\frac{r}{r_H} \right) \,,
\end{equation}
where $r_H$ is the location of the event horizon. 
The horizon is at $u=0$ limit, while for $u \rightarrow \infty$ the solution approaches AdS$_4$ spacetime, with radius $l_{AdS}=1/2$, which is kept fixed. Therefore, at the operative level we will work with the following metric:
\be
\label{eq:uMetric}
ds^2=- e^{2u-\beta(u)} \,\, r_H^2 \,\, H(u) \, dt^2 +\frac{du^2}{H(u)}+ e^{2u}\,\, r_H^2 d\Omega_2^2 
\,,
\qquad 
h(u) = r_H^2 \, e^{2u} \,H(u)\,.
\ee

The Reissner-Nordstr\"om “universal" solution is obtained by setting all the scalar fields to their the vacuum values \eqref{vacua_scalar} throughout the entire solution. The warp factors are:
\begin{align}
\beta &= 0 \ , &  H(u) &= 4+ \kappa \frac{e^{-2 u}}{r_H^2}- \frac{(16 r_H^4 + 4 \kappa r_H^2 + (P^1)^2 + Q_1^2) \, e^{-3u} }{4r_H^4} +\frac{((P^1)^2+Q_1^2) \, e^{-4 u}}{4 r_H^4}\,, \label{hrn}
\end{align}
with the following additional conditions coming from the scalar equations of motion:
\be
P^2 =0\,, \qquad Q_2=0\,.
\ee
The Reissner-Nordstr\"om solution in \eqref{hrn} is characterized by the two charges $Q_1$ and $P^1$, and the mass $M$ (alternatively, the radius of the horizon $r_H$).

\subsubsection{Strategy for numeric simulations}

To find the asymptotically AdS black hole and black brane solutions of interest, we will numerically solve the equations of motion subject to the appropriate boundary conditions. 

The equations of motion are given by the Euler--Lagrange equations derived from the action using the metric ansatz \eqref{eq:uMetric} as well as \eqref{scalarR}--\eqref{ans_vec} for the matter fields.
This leads to a system of coupled ODEs that determine the radial profile of the fields. 
The solutions depend on a number of integration constants, equal to the number of “degrees of freedom” one would expect from each of the fields.
For example, the scalar fields have a second order equation of motion, contributing 2 integration constants each.
The massless vector fields only contribute 1 each, the other one being constrained by a conservation law.

The boundary conditions are determined by physical requirements at the boundary of AdS and at the event horizon.
Asymptotically, we restrict the scalar fields to approach the value that minimizes the potential \eqref{eq:potential}, which provides the negative cosmological constant.
The heavy scalar fields and the massive vector have solutions that diverge near the boundary, the coefficients of which are required to vanish.
We will describe the resulting asymptotic boundary conditions in Sec. \ref{sec:asymptoticBehavior}.
Near the black hole horizon, where \( H \to 0 \), we require the spacetime to remain regular and the energy-momentum tensor to stay finite.
We will explain the result in Sec. \ref{sec:BHCond}.

These boundary conditions fix some of the integration constants in terms of the others.
For generic values of the free parameters at the boundary, \( \lambda^i_\text{bdy} \), the solution to the radial equations of motion will not satisfy the horizon boundary conditions, and vice versa.
To find solutions which satisfy both sets of boundary conditions simultaneously, we employ a shooting algorithm where we start from two solutions, one satisfying the asymptotic boundary conditions and parameterized by \( \lambda^i_\text{bdy} \), and another one satisfying the horizon boundary conditions and parameterized by \( \lambda^i_\text{BH} \). 
We then tune these parameters using a numerical minimization algorithm so that the fields and their derivatives match at an arbitrary radius between the boundary and the horizon.

\renewcommand{\arraystretch}{1.25}
\begin{table}[htbp]
	\centering
	\begin{tabular}{r | c c c c c c c c c c}
		field  &  H  &  \(\beta\)  &  \(\sigma\)  &  \(\tau_1\)  &  \(\tau_2\)  &  \(\tau_3\)  &  \(b_1\)  &  \(b_2\)  &  \(b_3\)  &  \(\zeta\)
		\\
		\hline
		horizon parameters  &  0  &  1  &  1  &  1  &  1  &  1  &  1  &  1  &  1  &  1
		\\
		boundary parameters  &  1  &  0  &  1  &  2  &  1  &  2  &  2  &  1  &  2  &  1
		\\
		matching conditions  &  -1  &  -1  &  -2  &  -2  &  -2  &  -2  &  -2  &  -2  &  -2  &  -2
		\\
		\hline
		\(Q^{111}\)  &  \(\times\)  &  \(\times\)  &  \(\times\)  &  \(\times\)  &  \(\times\)  &  \(\times\)  &  \(\times\)  &  \(\times\)  &  \(\times\)  &  \(\times\)  
		\\
		\(Q^{111}_e\)  &  \(\times\)  &  \(\times\)  &  \(\times\)  &  \(\times\)  &  \(\times\)  &  \(\times\)  &    &    &    &  \(\times\)  
		\\
		\(Q^{111}_m\)  &  \(\times\)  &  \(\times\)  &  \(\times\)  &  \(\times\)  &  \(\times\)  &  \(\times\)  &    &    &    &    
		\\
		\(M^{111}\)  &  \(\times\)  &  \(\times\)  &  \(\times\)  &  \(\times\)  &  \(\times\)  &  &  \(\times\)  &  \(\times\)  &  &  \(\times\)  
	\end{tabular}
	\caption{Overview of the degrees of freedom for the boundary value problem in each model. The horizon and boundary parameters are discussed in \Cref{sec:BHCond,sec:asymptoticBehavior}, respectively. The number of matching conditions for each field is given by the order of its equation of motion. This is usually referred to as the number of (dynamical) degrees of freedom, but they count negatively towards the number of free parameters of the boundary value problem.}
	\label{tab:dofOverview}
\end{table}

For each field in the \(Q^{111}\) model, the number of free parameters at the black hole horizon and at the boundary of AdS is given in \Cref{tab:dofOverview}, as well as the number of conditions that must be matched in the middle.%
\footnote{The counting at the AdS boundary must be taken with a grain of salt. The behavior near the conformal boundary is characterized by the fields that diagonalize the mass matrix. These are nontrivial combinations of the fields in \Cref{tab:dofOverview}. So it is not quite correct to assign boundary parameters to the fields in this table. Nevertheless, the total number of free boundary parameters in \Cref{tab:dofCounting} is correct for each of these models.}
We also indicate the field content that remains in the \(M^{111}\) model, as well as the models obtained by restricting to purely electric ($Q^{111}_e$) or purely magnetic ($Q^{111}_m$) black holes. This is a further simplification that we will adopt later, as it allows us to consistently set the axions $b^i$ to zero, in which case the period matrix $\mathcal{N}_{\Lambda\Sigma}$ is purely imaginary.

To get the final counting of free parameters, the electromagnetic charges and (for spherical and hyperbolic black holes) the black hole radius need to be taken into account, as in \Cref{tab:dofCounting}. This gives the expected dimensionality of the solution space. In principle, this counting could be off, due to the nonlinearity of the system, but in practice we find that it is correct.

\begin{table}[htbp]
	\centering
	\begin{tabular}{r | c c c c c | c}
		&  horizon  &  boundary  &  matching  &  charges  &  radius  &  total
		\\
		\hline
		\(Q^{111}\)  &  9  &  13  &  -18  &  6  &  1 (0)  &  \textbf{11} (10)
		\\
		\(Q^{111}_e\)  &  6  &  8  &  -12  &  3  &  1 (0)  &  \textbf{6} (5)
		\\
		\(Q^{111}_m\)  &  5  &  7  &  -10  &  3  &  1 (0)  &  \textbf{6} (5)
		\\
		\(M^{111}\)  &  7  &  9  &  -14  &  4  &  1 (0)  &  \textbf{7} (6)
	\end{tabular}
	\caption{Counting the number of free parameters in each model, using the degrees of freedom in \Cref{tab:dofOverview} and the charges. The radius of the black hole is a free parameter for spherical black holes and hyperbolic black branes. For planar black branes, there is a symmetry\textsuperscript{\ref{fn:planarSymmetry}} that effectively removes this degree of freedom, reducing the dimensionality of the solution space as indicated in brackets.}
	\label{tab:dofCounting}
\end{table}

\subsubsection{Asymptotic fall off of the fields}
\label{sec:asymptoticBehavior}

Near the AdS boundary, the solutions have a characteristic fall-off, polynomial (or logarithmic) in terms of the Fefferman--Graham coordinate $z = 1/r = e^{-u}/r_H$. 
We can obtain the boundary conditions by solving the equations of motion perturbatively in \( z \).
At each order in \( z^n \log(z)^m \), with \( n, m > 0 \) to keep the stress tensor finite, the equations of motion determine all but a few coefficients in the Taylor expansion of the fields.
The coefficients left undetermined are the parameters in \Cref{tab:dofOverview} that distinguish the different solutions.

As an illustration, consider the scalar fields \( \tau_i \) with expansion \( \tau_i(z) = 1 + \tau_i^{(1)} z + \mathcal{O}(z^2) \).
The first term ensures that the potential \eqref{vacua_scalar} is minimized asymptotically. 
The scalar equations of motion further imply, at linear order in \(z\), that \( \tau_1^{(1)} + \tau_2^{(1)} + \tau_3^{(1)} = 0 \).
We take this constraint to fix \( \tau_3^{(1)} \) in terms of \( \tau_1^{(1)} \) and \( \tau_2^{(1)} \). The later remain unconstrained by the equations of motion at any order.

Performing this process systematically for all fields and equations of motion up to 6th order in \(z\) (indeed, the most massive field has \(\Delta = 6\)) we find that in the \(Q^{111}\) compactification, all coefficients can be determined in terms of the charges and 13 parameters 
\begin{align}
\label{eq:coeffs}
	(\tau_1^{(1)}, \tau_2^{(1)}, b_1^{(1)}, b_2^{(1)}, \tau_1^{(2)}, \tau_1^{(2)}, b_1^{(2)}, b_2^{(2)}, H^{(3)}, \sigma^{(4)}, \zeta^{(4)}, b_1^{(5)}, \sigma^{(6)} ) \ .
\end{align}
They parameterize the following asymptotic behavior.
\begin{subequations} \label{exp_infinity}%
	The components of the metric are
	\begin{align}
	\label{eq:metricFallOff}
		H
		&= 4 + \left(\kappa + 2 \tau_1^{(1) 2} + 2 \tau_2^{(1) 2} + 2 \tau_1^{(1)} \tau_2^{(1)} + 2 b_1^{(1)2} + 2 b_2^{(1) 2} + 2 b_2^{(1)} b_2^{(1)} \right) z^2 + h_{(3)} z^3 + \order{z^4} \nonumber \\
		\beta
		& = \frac12 \left( \tau_1^{(1) 2} + \tau_2^{(1) 2} + \tau_1^{(1)} \tau_2^{(1)} + b_1^{(1) 2} + b_2^{(1) 2} + b_1^{(1)} b_2^{(1)} \right) z^2 + \order{z^3}
		\ ,
	\end{align}
	The AdS-Reissner--Nordström solution with $M = -h_{(3)} / 2$, as in \eqref{hrn}, is a solution for which all coefficients in \eqref{eq:coeffs} except for $H^{(3)}$ vanish.
	As mentioned before, we choose the time coordinate such that $\beta|_{z=0} = 0$.
	
	The scalar fields have the following expansion as $z \to 0$ (for the sake of clarity, we omit terms that are at least quadratic in the coefficients)
	\begin{align}
	\label{eq:scalarExpansion}
		\tau_1
		& = 1 + \tau_1^{(1)} z + \tau_1^{(2)} z^2 + \ldots + \left( \frac43 \sigma^{(4)} - \frac{\kappa}{12} \tau_1^{(2)} + \ldots \right) z^4 + \ldots
		\nonumber \\
		&\pheq - \left( \sigma^{(6)} + \frac{\kappa}2 \sigma^{(4)} - \frac{\kappa^2}{80} \tau_1^{(2)} + \ldots \right) z^6 + \order{z^7}
		\nonumber \\
		\tau_2
		& = 1 + \tau_2^{(1)} z + \tau_2^{(2)} z^2 + \ldots + \left( \frac43 \sigma^{(4)} - \frac{\kappa}{12} \tau_2^{(2)} + \ldots \right) z^4 + \ldots
		\nonumber \\
		&\pheq - \left( \sigma^{(6)} + \frac{\kappa}2 \sigma^{(4)} - \frac{\kappa^2}{80} \tau_2^{(2)} + \ldots \right) z^6 + \order{z^7}
		\displaybreak[0] \nonumber \\
		\tau_3
		&= 1 - (\tau_1^{(1)} + \tau_2^{(1)}) z - \left( \tau_1^{(2)} + \tau_2^{(2)} + \ldots \right) z^2 + \ldots + \left( \frac43 \sigma^{(4)} + \frac{\kappa}{12} (\tau_1^{(2)} + \tau_2^{(2)}) + \ldots \right) z^4
		\nonumber \\
		&\pheq + \ldots - \left( \sigma^{(6)} + \frac{\kappa}2 \sigma^{(4)} + \frac{\kappa^2}{80} (\tau_1^{(2)} + \tau_2^{(2)}) + \ldots \right) z^6 + \order{z^7}
		\nonumber \displaybreak[0] \\
		b_1
		& = b_1^{(1)} z + b_1^{(2)} z^2 + \ldots - \left( \frac{\kappa}{12} b_1^{(2)} + \ldots \right) z^4 + (b_1^{(5)} + \ldots) z^5 + \left( \frac{\kappa^2}{80} b_1^{(2)} + \ldots \right) z^6 + \order{z^7} \nonumber \displaybreak[0] \\
		b_2
		& = b_2^{(1)} z + b_2^{(2)} z^2 + \ldots - \left( \frac{\kappa}{12} b_2^{(2)} + \ldots \right) z^4 + (b_1^{(5)} + \ldots) z^5 + \left( \frac{\kappa^2}{80} b_2^{(2)} + \ldots \right) z^6 + \order{z^7} \nonumber \displaybreak[0] \\
		b_3
		& = - (b_1^{(1)} + b_2^{(1)}) z - (b_1^{(2)} + b_2^{(2)} + \ldots) z^2 + \ldots + \left( \frac{\kappa}{12} (b_1^{(2)} + b_2^{(2)}) + \ldots \right) z^4
		\nonumber \\
		&\pheq + \left( b_1^{(5)} + \ldots \right) z^5 - \left( \frac{\kappa^2}{80} (b_1^{(2)} + b_2^{(2)}) \right) z^6 + \order{z^7} 
		\nonumber \\
		\sigma &= 1 + \ldots + \left( \sigma^{(4)} + \ldots \right) z^4 + \ldots + \left( \kappa \sigma^{(6)} + \ldots \right) z^6 + \order{z^7}
	\end{align}
	This asymptotic behavior is consistent with the masses $m^2 l_{AdS}^2 = (18, 10, 4, -2, -2, -2, -2)$ that we found in \cref{sec:consistentTruncation}: there are two independent components for each of the fields with mass $-2$, and one for each field that is more massive.
	Interactions give rise to terms quadratic and higher order in these coefficients or in the charges, which are included in the ``$\ldots$''.
	
	The massive vector field $\zeta$ falls off as
	\begin{align}
	\label{eq:vectorExpansion}
		\zeta 
		&= \ldots + \left( \zeta^{(4)} + \ldots \right) z^4 - \left( \frac16 \zeta^{(4)} + \ldots \right) z^6 + \order{z^7} \,.
	\end{align}
	This asymptotic behavior \( z^4 \) is compatible with the mass of the vector field and is related to the expectation value of a dual operator with scaling dimension $\Delta=5$.
	Nevertheless, interactions with the other fields give rise to terms proportional to the conserved charges that fall off more slowly near the boundary.
	
\end{subequations}

\subsubsection{Boundary conditions at the horizon}
\label{sec:BHCond}

The event horizon is located where \( g_{tt} \propto H \) vanishes.
We choose our coordinates so that this happens at \( u = 0 \).
In order to ensure the regularity of spacetime, the Einstein equations impose that the energy momentum tensor must remain finite.
This would not be the case if the massive vector field \(\zeta\) had a nonzero value at the horizon.
Using the equations of motion, we find that the behavior of the fields at the horizon is characterized by 9 parameters for the \(Q^{111}\) model
\begin{align}
	\left( \beta^{(h)}, \sigma^{(h)}, \tau_1^{(h)}, \tau_2^{(h)}, \tau_3^{(h)}, b_1^{(h)}, b_2^{(h)}, b_3^{(h)}, \zeta'^{(h)} \right)
	\ ,
\end{align}
which determine the values of the fields in the obvious way
\begin{align}
	\beta &= \beta^{(h)} + \order{u}
	\ , &
	\sigma &= \sigma^{(h)} + \order{u}
	\ , &
	\zeta &= \zeta'^{(h)} u + \order{u^2}
	\ ,
	\nonumber \\
	\tau_1 &= \tau_1^{(h)} + \order{u}
	\ , &
	\tau_2 &= \tau_2^{(h)} + \order{u}
	\ , &
	\tau_3 &= \tau_3^{(h)} + \order{u}
	\ ,
	\nonumber \\
	b_1 &= b_1^{(h)} + \order{u}
	\ , &
	b_2 &= b_2^{(h)} + \order{u}
	\ , &
	b_3 &= b_3^{(h)} + \order{u}
	\ .
\end{align}
The higher order terms are determined by the equations of motion.
For example, the first derivative of \( H \) near the horizon is
\begin{align}
\label{eq:Hprime}
	H'(0) &= 12 - 2 
	\begin{aligned}[t]
		&\left[ 24 \tilde{\sigma}^{(h)} - 8 \tilde{\sigma}^{(h)} (\tilde{\tau}_1^{(h)} + \tilde{\tau}_2^{(h)} + \tilde{\tau}_3^{(h)}) \right.
		\nonumber \\
		&\pheq + \tilde{\tau}_1^{(h) 2} + \tilde{\tau}_2^{(h) 2} + \tilde{\tau}_3^{(h) 2} + 4(\tilde{\tau}_1^{(h)} \tilde{\tau}_2^{(h)} + \tilde{\tau}_2^{(h)} \tilde{\tau}_3^{(h)} + \tilde{\tau}_1^{(h)} \tilde{\tau}_3^{(h)})
		\nonumber \\
		&\pheq + \left. \tilde{b}_1^{(h) 2} + \tilde{b}_2^{(h) 2} + \tilde{b}_3^{(h) 2} + 4 (\tilde{b}_1^{(h)} \tilde{b}_2^{(h)} +  \tilde{b}_2^{(h)} \tilde{b}_3^{(h)} + \tilde{b}_1^{(h)} \tilde{b}_3^{(h)}) \right]
	\end{aligned}
	\nonumber \\
	&\pheq + \frac{\kappa - \tfrac14 \zeta^{(h)}}{r_H^2} - \frac{P_1^2 + P_2^2 + 2 P_3^2 + p_{\xi_1}^2 + p_{\xi_2}^2 + \tfrac12 p_{\xi_3}^2}{4 r_H^4} + \ldots
	\ ,
\end{align}
where we have included only terms that are at most quadratic in the difference between the horizon coefficients and the asymptotic values, i.e.~we include terms quadratic in the coefficients \( (\beta^{(h)}, b_1^{(h)}, b_2^{(h)}, b_3^{(h)}, \zeta^{(h)}) \) as well as in \( (\tilde{\sigma}^{(h)} \equiv \sigma^{(h)} - 1, \tilde{\tau}_1^{(h)} \equiv \tau_1^{(h)} - 1, \tilde{\tau}_2^{(h)} \equiv \tau-2^{(h)} - 1, \tilde{\tau}_3^{(h)} \equiv \tau_3^{(h)} - 1) \).

This explains the counting in \Cref{tab:dofCounting} for the \(Q^{111}\) compactification. The other models can be obtained by further constraining certain coefficients at outlined at the end of Sec. \ref{sec:consistentTruncation}. 
In the following sections we will give several examples of black hole solutions that we have found with this shooting technique.

\subsubsection{Thermodynamics and stability}
  
The solutions we have found are characterized by temperature
  \be
  T = \frac{e^{-\beta/2} h'}{4 \pi} \bigg|_{r=r_H} \ ,
  \ee
  which can be rewritten in terms of \cref{eq:Hprime} using that \( h' = r_H^2 H' \) on the horizon.
  The other thermodynamic quantities such as the mass and the entropy are
  \be
  M = -\frac1{12} \partial_z^3 H \bigg|_{z = 0} \,,  \qquad S = \pi r_H^2 \,,
  \ee
  and the free energy can be found by suitably renormalizing the on-shell action, see \cite{Monten:2016tpu}, where the thermodynamics of the spherical solutions was studied. It was shown that a liquid-gas-like first order phase transition arises between small and large black holes, once the charges are below a certain critical value. Across the phase transition, the black hole “swallows" the massive vector halo, and the process turns a small black hole with a larger value for the massive vector field into a black hole with large area and small value for the massive vector. See \cite{Monten:2016tpu}, Sec. 5 for further details.

We will now turn to the study of the possibility of brane nucleation from various black holes.
We will work in the canonical ensemble of fixed temperature and electromagnetic charges: unless stated otherwise, all the solutions we use as background have positive specific heat
\be
C_Q = T \left( \frac{d S}{ d T}\right)_Q \,,
\ee
namely the (single center) background solutions per se are thermodynamically stable. 

We will see that in certain cases they are unstable due to the possibility of nucleating various kinds of branes. In what follows we will work with purely electric or purely magnetic backgrounds, so that the solutions themselves are supported by scalars which have no axions. We will probe these backgrounds against different kinds of instabilities, in particular nucleation of wrapped probe branes and spacetime filling ones. The former will appear point-like in the four-dimensional AdS spacetime and form real black hole bound states. We turn to the latter in the next section.

\section{Spacetime filling probe branes \label{spacetime_filling}}

As a first example of instability towards brane nucleation, in this section will analyze the (in)stability towards nucleating spacetime filling M2 branes of the (purely electric and purely magnetic) planar solutions that we can find with the methods outlined above. These branes are of the same type as those which “provide" the dual field theory, a negative minimum outside the horizon signals an instability that leads to a Higgsing of the gauge group, in other words there is a spontaneous breaking of the gauge symmetry. 

\subsection{Spacetime filling M2 brane}

We consider here configurations of probe M2 branes which extend along the $t,x,y$ spacetime components, in Poincar\'e coordinates ($\kappa = 0$). The action for a probe M2 brane in our black brane background is
\be \label{SM2}
S_{M2} = - \tau_{M_2} \int d^3x \sqrt{G} \pm \tau_{M_2} \int C_3 = -\tau_{M_2} \int dt dx^1 dx^2 (V_g + V_e) \,.
\ee
We will consider backgrounds with nonzero value for the vector field $A^0$. In the M-theory uplift of Sec. \ref{upliftfomulae} one can see that the R-charge corresponds to angular momentum: the black brane in the eleven dimensional spacetime is rotating in the $\psi$ direction. Let us parameterize the coordinates of the M2 branes with $(\tau, \chi^1,\chi^2)$, all noncompact, ranging from $- \infty$ to $+\infty$. This section follows \cite{Henriksson:2019ifu}, where the possibility of D3 brane nucleation on black branes from the $T^{11}$ truncation is studied. We will stick to their notation to facilitate comparison between the two cases. We proceed by making the following ansatz for the embedding:
\be
\Theta_1 = \theta_0^1\,,  \qquad \Phi_1 = \phi_0^1\,, \qquad \Theta_2 = \theta_0^2\,,  \qquad \Phi_2 = \phi_0^2\,, \qquad \Theta_3 = \theta_0^3\,,  \qquad \Phi_3 = \phi_0^3\,,
\ee
\begin{equation*}
\mathcal{T} = \mathcal{T}(\tau)\,, \qquad \mathcal{R} = \mathcal{R}(\tau)\,, \qquad X_i = \chi_i \,, \,\,\,(i =1,2) \qquad \Psi = \Psi(\tau)\,,
\end{equation*}
where the quantities with the subscript $_0$ denote the constant position of the brane in the internal coordinates. Consider an observer located on the brane at fixed worldvolume coordinates. Let the velocity be
\be
U \equiv \frac{d X^{\mu}}{d \tau} \partial_\mu = \dot{\mathcal{T}} \partial_t + \dot{\mathcal{R}} \partial_r+ \dot{\mathcal{\psi}} \partial_{\psi}\,,
\ee
so that 
\be
U_{\mu} U^{\mu} = g_{tt} \dot{\mathcal{T}}^2 + g_{rr} \dot{\mathcal{R}}^2 + g_{\psi \psi}  \dot{\psi}^2 + 2 g_{t \psi} \dot{\mathcal{T}} \dot{\psi}
\ee
will eventually be set to $-1$ by an appropriate choice of worldvolume time $\tau$.
The induced line element is then
\be
ds^2_3 = U_{\mu} U^{\mu} d \tau^2 + g_{xx} (d \chi_1^2 + d\chi_2^2)\,.
\ee
The gravitational part of the probe action \eqref{SM2} reads: 
\be
\sqrt{- det (P[g_{\mu \nu}])} = g_{xx} \sqrt{- U_{\mu} U^{\mu}} \,,
\ee
whereas the WZ term of the action becomes
\be
P[C_3]  =  [ (C_3)_t \dot{\mathcal{T}} + (C_3)_{\psi} \dot{\psi} ] \, d\tau \wedge d\chi_1 \wedge d\chi_2\,,
\ee
where we have defined $(C_3)_t$ and $(C_3)_{\psi}$ as the $t, x,y $ and $\psi, x,y$ components of the M-theory 3-form.

We can now compute  the energy and the angular momentum of the M2 brane, varying the Lagrangian density with respect to $\dot{\mathcal{T}}$ and $\dot{\psi}$, arriving at the expressions:
\be
 E = -\frac{1}{\tau_{M_2}} \frac{\partial \mathcal{L}_{M_2}} {\partial \dot{\mathcal{T}}} = g_{xx} (-g_{tt} \dot{\mathcal{T}} - g_{t \psi} \dot{\psi}) - (C_3)_{t}\,,
\ee
\be
J=  \frac{1}{\tau_{M_2}} \frac{\partial \mathcal{L}_{M_2}} {\partial \dot{\mathcal{\psi}}} = g_{xx} (g_{t\psi} \dot{\mathcal{T}} + g_{\psi \psi} \dot{\psi})+ (C_3)_{\psi}\,,
\ee
where we have set $U_{\mu} U^{\mu} = -1$ after doing the variation. We can now express the energy in function of the angular momentum. In doing so we first choose the branch with $\dot{\mathcal{T}}>0$ when solving for $\dot{\mathcal{T}}$ from $U_{\mu} U^{\mu} = -1$. We arrive at 
\be
E= - (C_3)_{t} -g_{t\psi} \frac{J - (C_3)_{\psi}}{g_{\psi \psi}} + \frac{\sqrt{g_{t\psi}^2 - g_{\psi \psi} g_{tt}} \sqrt{((C_3)_{\psi}^2 - J)^2 + g_{\psi \psi} g_{xx}^2 (1 + g_{rr} \dot{\mathcal{R}}^2) }}{g_{\psi \psi}}\,,
\ee
and we set $\dot{\mathcal{R}} =0$ in order to extract the form of the effective potential for the probe M2 brane:
\be
V_{M_2} = - (C_3)_{t} -g_{t\psi} \frac{J - (C_3)_{\psi}}{g_{\psi \psi}} + \frac{\sqrt{g_{t\psi}^2 - g_{\psi \psi} g_{tt}} \sqrt{(C_3)_{\psi}^2 -2 (C_3)_{\psi} J + J^2 + g_{\psi \psi} g_{xx}^2}}{ g_{\psi \psi}}\,.
\ee
After we plug in the explicit metric components, we obtain
\be \label{effective_spacetime_filling}
V_{M2} = - (C_3)_{t} - A_t^0(r) ( J - (C_3)_{\psi} ) +  \frac{\sqrt{ h} e^{-\beta/2}}{\sqrt{\tau_1 \tau_2 \tau_3}}  \sqrt{ ((C_3)_{\psi}-J)^2 + \sigma^{4} r^4}  
\ee
where $A^0$ is defined in \eqref{eigen_vec}.  Moreover $Q_R =Q_0 = J$ is the electric charge  associated  to $A^0$. The components of the 3-form can be inferred from \eqref{eq:fourForm},
\begin{eqnarray}
(C_3)_t & = & \int dr \, \left( 6 e^{-\beta/2} r^2 \,   \mathcal{K} ^{-1} \sigma^{4}  + 4 \sqrt3  ( e^{-\beta/2}r^2 \mathcal{B}_t) A^0 (r) \sigma^{4} \right) = \nonumber \\
& = & \int dr \left[ 6 \left(\frac{\sigma^{4} }{\tau_1 \tau_2 \tau_3} \right) r^2 \, e^{-\beta/2} + 4 \sqrt3  ( e^{-\beta/2}r^2 \mathcal{B}_t) A^0 (r) \sigma^{4} \right]\,, \nonumber
\\
\qquad (C_3)_{\psi} & = & 4 \sqrt3  \int dr  e^{-\beta/2}r^2 \mathcal{B}_t \sigma^{4} \,.
\end{eqnarray}

We are now going to plot the potential for M2 branes in different examples, with and without the R-symmetry field $A^0$, whose associated charge is interpreted as the angular momentum for the probe branes.

\subsubsection{Purely magnetic and purely electric background black branes}

As a warm up case, we first identify the two Betti vectors and switch off the electric charges. This means that the solution, written in our symplectic frame will only have $P_2$ charge. Therefore, $A^0 = 0$, $(C_3)_{\psi} =0 $ and $J_{\psi} =0$, so that the probe potential reduces to:
\be
V_{M2} = - (C_3)_{t}  +  \sqrt{h} e^{-\beta /2} \frac{\sigma^2}{\sqrt{\tau_1 \tau_2 \tau_3}}  r^2 \,, \ee
which, upon the change of variables
\be
\frac12 \log(\tau_1) = \frac{3 \chi -2 \eta_2 -2 \eta_3}{4} \,,
\qquad 
\frac12 \log(\tau_2) = \frac{3 \chi -2 \eta_1 -2 \eta_3}{4} \,,
\qquad 
\frac12 \log(\tau_3) = \frac{3 \chi -2  \eta_2 -2 \eta_1}{4} \,,
\ee
and $
\phi = \frac{-3 \chi - \eta_1 -\eta_2 - \eta_3}{2} \label{phitoeta}
$
coincides with the potential found in \cite{Klebanov:2010tj}\footnote{Recall that in our conventions $e_0=6$ and $l_{AdS}^2=1/4$, and $\beta = w$.}:
\be
V_g  (r)= \sqrt{G} =  e^{-21\chi/4} \sqrt{g}e^{-\beta/2} r^2\,,
\qquad
V_e' (r) = \pm 6  r^2 e^{-\beta/2-21\chi/2}\,.
\ee
The $-$ sign corresponds to an M2-brane of the same type as those sourcing the background (the plus sign instead corresponds to a brane of opposite charge, which is always attracted to the black brane). We will work with the minus sign. In \cite{Klebanov:2010tj} it was found that the potential increases monotonically and never dips below the horizon  value,  so  the  background  is  stable  with  respect  to  tunneling  of  spacetime  filling M2-branes. We reproduced their result as a control case, see \Cref{region_insta}.
    
\begin{figure}[H]
\begin{center}
    \includegraphics[width=130mm]{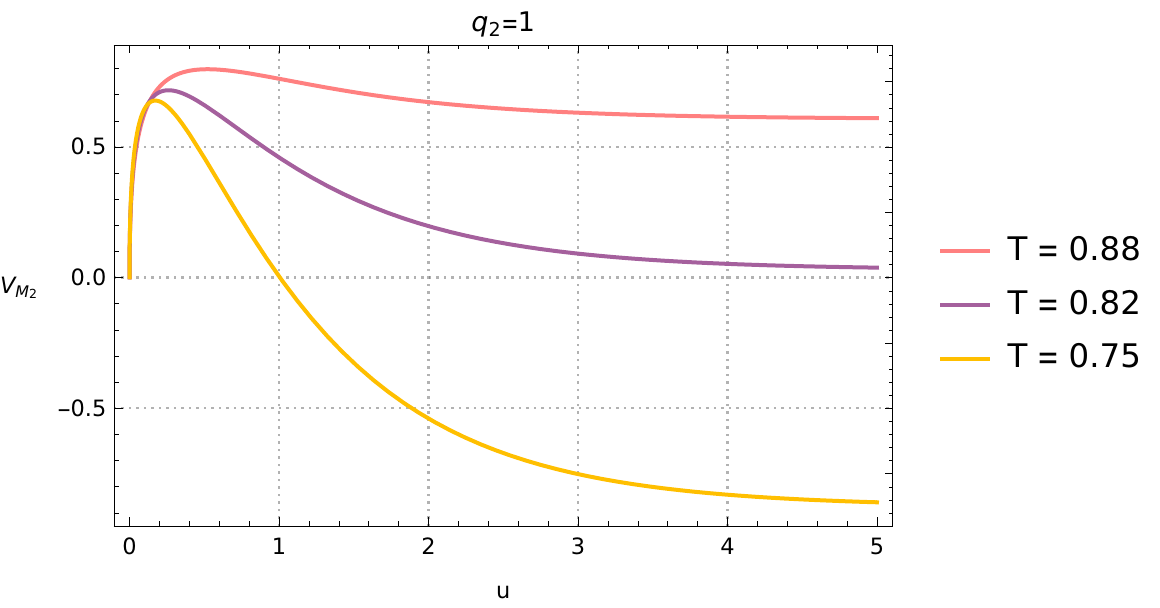}
    \end{center}
    \caption{M2 spacetime filling brane potential for the planar AdS$_4$ solution with R-symmetry field turned on. The value of $Q_2$ is kept fixed, and the plot displays solutions with $Q_1 = -0.1$ (pink), $Q_1 = 0.9$ (violet) and $Q_1 = 1.5$ (yellow). \label{planar_st_nucleation}}
  \end{figure}

\begin{figure}[H]
\begin{center}
    \includegraphics[width=77.5mm]{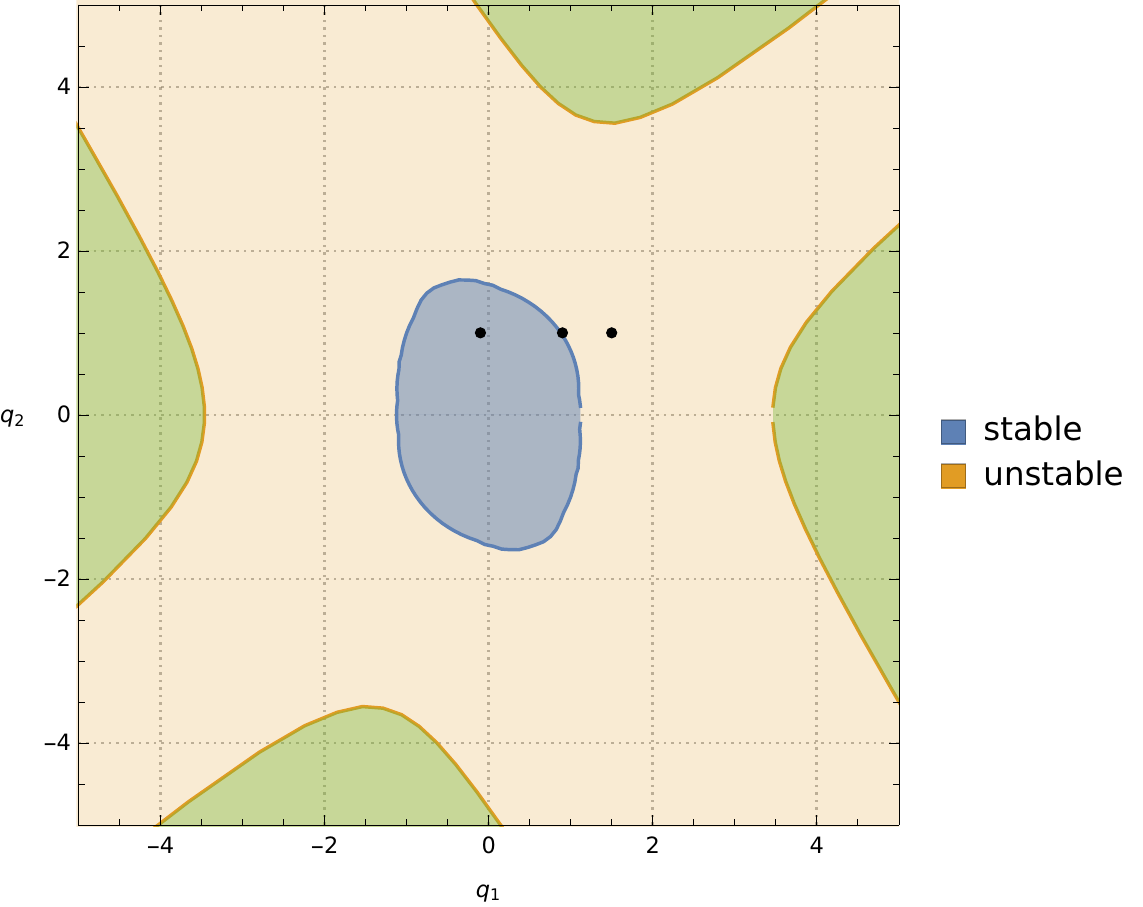}
        \includegraphics[width=77.5mm]{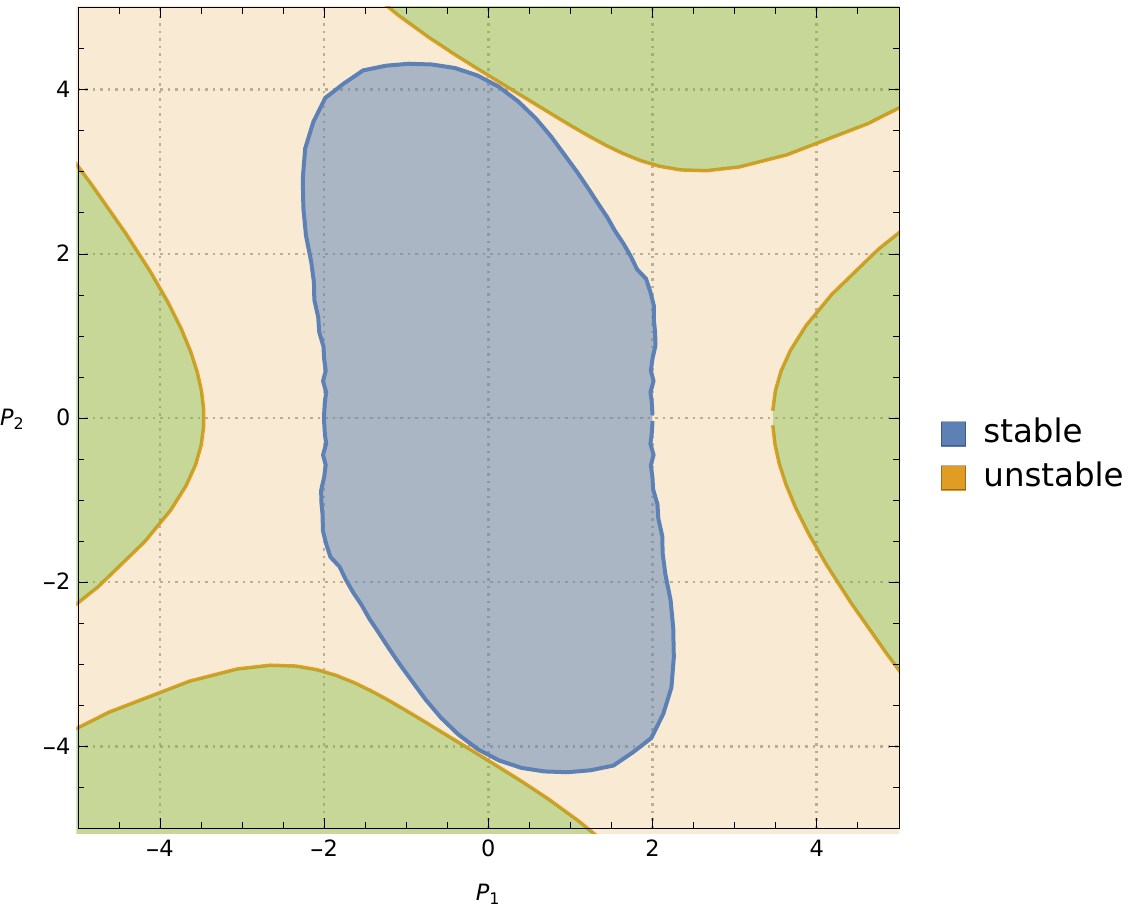}
    \end{center}
    \caption{In blue, the region of parameter space where the black branes are stable with respect to to nucleation of spacetime filling M2 branes. The orange region contains unstable solutions. No black brane solutions exist in the green area, as they would be over-extremal. The graph on the left refers to purely electric black branes (the black dots refer to the three solutions plotted in Fig. \ref{planar_st_nucleation}), the one on the right to purely magnetic ones, both with R-symmetry gauge field turned on. The stability result of \cite{Klebanov:2010tj} corresponds to the vanishing unstable region on the $P_2$ axis (for $P_1 = 0$), where the blue and green regions touch. 
    \label{region_insta}}
  \end{figure}

The situation becomes more interesting if we consider solutions with nonzero R-symmetry charges $P_1$ or $Q_1$. Indeed the effective potential \eqref{effective_spacetime_filling}, which is normalized to be zero at the horizon, asymptotically approaches a negative value for sufficiently low temperature in, for example, purely electric solutions - see Fig. \ref{planar_st_nucleation}. This means that it is energetically favorable for a black brane to nucleate probe M2 branes that move away towards the boundary of spacetime, confirming the fact that these sort of horizons suffer from an instability close to extremality \footnote{Another instability for magnetic branes in $S^7$ truncations of M-theory as found in \cite{Donos:2011bh,Donos:2011qt}. The latter instability is of a different nature: it concerns spatially modulated instability and violations of the BF bound in the near horizon AdS$_2$ region. It will be nevertheless interesting to investigate whether there is a relation between this and the instability we have found.}. We find the same behaviour when we consider purely magnetic branes: also in this case, a nonzero (magnetic) component of the R-symmetry field triggers the instability towards M2 brane nucleation (the massive vector field is switched off in this case). In Fig. \ref{region_insta} we show the regions in phase space where the instability is present, for the electric (left) and magnetic (right) cases.

In a way, the kind of potential displayed in Fig. \ref{planar_st_nucleation} does not come as a surprise given that the R-symmetry field is turned on, and the brane is effectively rotating in 11d: the probe branes experience a centrifugal force that pushes them away from the center. In this case the dimension of the M-theory direction (the function $e^{2V}$ in \eqref{e2v}) decreases near the BH for all solutions, the decrease being sharper for the solutions with the instability towards M2 decay.

To end this section, let us mention that a similar instability driven by R-symmetry field was found in the case of spacetime-filling D3 branes in conifold truncations \cite{Henriksson:2019ifu}, and was interpreted as a color superconductor instability in which the rank of the group is changed with the expulsion of branes.

\section{Wrapped probe branes \label{wrapped}}

In this section we focus on the stability of probe branes  wrapped around non-contractible cycles of the internal $Q^{111}$ and $M^{111}$ manifold. We first analyze wrapped M2 branes: for this case we find that there are no stable probes outside the horizon of the simple class of purely magnetic solutions with spherical horizons. Next, we perform the formal reduction to type IIA and turn our attention to the stability of probe fluxed D6 branes, which correspond to states of KK-monopoles, M5, M2 branes and KK modes in the M-theory picture. These latter possess a combination of charges which is more likely to find bound states \cite{Anninos:2013mfa}, given our background solutions. We find that stable fluxed D6 probes exist on purely electric backgrounds, confirming the possible presence of multicenter black holes in this sort of truncations. We also consider magnetic backgrounds that admit a supersymmetric limit, however we find that a local minimum of the effective potential does not form in this regime.

\subsection{Wrapped M2 branes}
In this section we consider for simplicity the same setup described in \cite{Klebanov:2010tj}, namely solutions with $A^0=0$ and purely baryonic charges\footnote{This choice allows us to borrow the results of \cite{Klebanov:2010tj} for the minimal volume cycles. A non-zero value for $A^0$ could in principle alter the computation.}. Our solutions will however have spherical horizons, while in \cite{Klebanov:2010tj} only planar horizons were considered. We start from the action
\be
S_{M2}^{\mathcal{C}} = - \tau_{M_2} \int d^3x \sqrt{-det(g)} + \tau_{M_2} \int C_3\,,
\ee
where $\tau_{M_2}$ is the M2 brane tension, and we again split it in two parts
\be
S_{Dp} = S_{grav}+S_F \,.
\ee
When we wrap M2 brane on one of the 2-cycles inside $Q^{111}$, the gravitational potential is proportional to the volume of such cycles, while $S_F$ is topological: minimizing the volume increases the possibility of finding stable probes. A  simple  way  to  construct  non-trivial  two-cycles  in the 7d manifold $Y^7$ is  to  start  with a 2-cycle in the six-dimensional base $B_6$ and lift it to $Y^7$. There are of course conditions that need to be satisfied that allow one to lift these 2 cycles on $Y^7$ and these were conveniently worked out in \cite{Klebanov:2010tj}, for a particular sub-truncation which corresponds in our conventions to zero R-symmetry field, purely magnetic configurations and hence vanishing massive vector and axions. The upshot of their analysis is that the cycle $\mathcal{C}$ should satisfy
\be
\int_{\mathcal{C}} J = 0\,,  \qquad \quad J = \omega_1 + \omega_2  + \omega_3\,,
\ee
with $\omega_i  $, $i =1,2,3$ defined in \eqref{omegas}.

\begin{figure}[H]
\begin{center}
    \includegraphics[width=130mm]{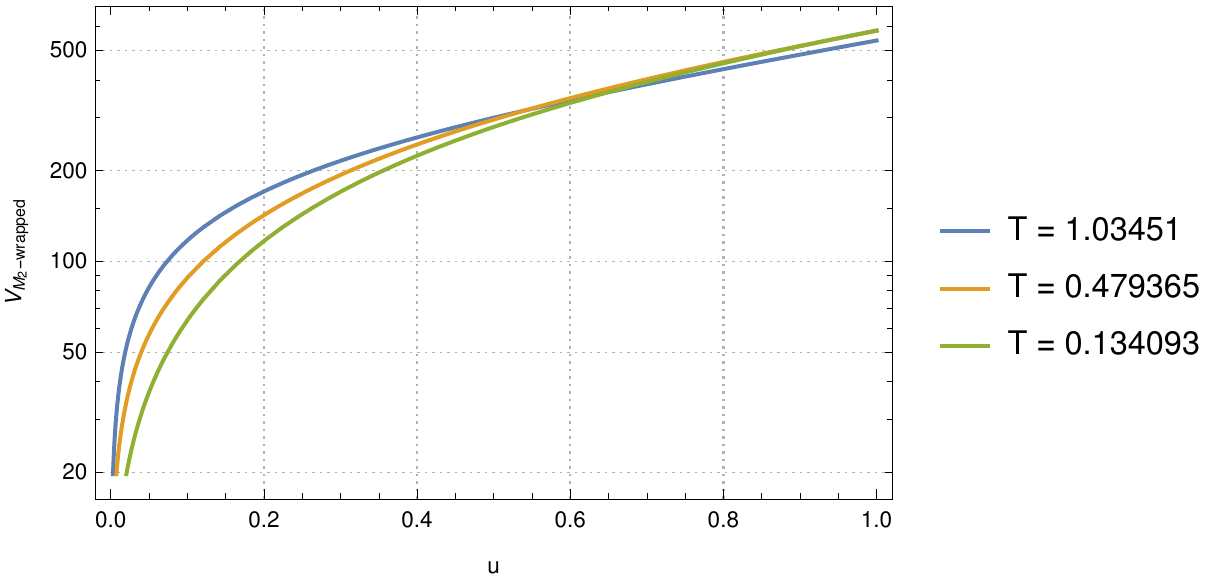}
    \end{center}
    \caption{Potential for M2 brane wrapped on $\mathcal{C}_2$  for the spherical AdS$_4$ solution without R-symmetry field, for configurations with magnetic charge $P_2 = 1/100$ (blue), $P_2 = 152/25$ (orange) and $P_2 = 8$ (green). \label{M2wrapped}}
  \end{figure}

We will focus our attention on the minimal volume cycle denoted as $\mathcal{C}_2$ in  \cite{Klebanov:2010tj}  \be
\mathcal{C}_2 : \left\{ \begin{array} {cccc}
\theta_1 & = & \theta_3 \\
\theta_2 & =  & const \\
\phi_1 & = & - \phi_3 \\
\phi_2 & = & const 
\end{array} \right.
\ee
We can now compute the gravitational and electrostatic potentials of M2 branes wrapped on $\mathcal{C}_2$. We choose to identify both Betti vectors as in \cite{Klebanov:2010tj}. Notice that the 4-form pulled back onto the M2 brane worldvolume in our solutions is 
\be
F_4 =  H_2^i \wedge \omega_i \,,
\ee
hence the electromagnetic potential reads:
\be
S_F = 4 \pi \varphi \,, \qquad 
\varphi'  = \frac{d \varphi}{ d r} = \frac14 \frac{8}{\sqrt6} \frac{P_2}{r^2} e^{-\beta/2} \left(\frac{1}{\tau_2} + 2 \frac{\tau_2}{\tau_1^2} \right)\,.
\ee
For the gravitational part we have \cite{Klebanov:2010tj}
\be
\sqrt{det}_{2} =   \sqrt{g} e^{-\beta/2} \left(\frac{1}{\sqrt{\tau_2}} + \frac{\sqrt{\tau_2}}{\tau_1} \right)\,.
\ee
So in total for the $\mathcal{C}_2$ in $Q^{111}$ we have:
\be
S_{Q^{111}}^{\mathcal{C}_2} = 4 \pi \left( \sqrt{g} e^{-\beta/2} \left(\frac{1}{\sqrt{\tau_2}} + \frac{\sqrt{\tau_2}}{\tau_1} \right) - \varphi  \right)\,.
\ee
The behaviour of the potential is depicted in Fig. \ref{M2wrapped} which shows that the probes are overwhelmingly gravitationally attracted by the black hole (no stable and metastable probes). 

To sum up, in this section we have found that wrapped M2 probe branes on black hole background solutions of the same nature (i.e. with only charges due to the Betti vector) cannot be stable. Consistently with the expectations from black hole bound states in asymptotically flat spacetimes, we expect that only mutually nonlocal charges can form a bound state. This is confirmed by our findings in the next section.

\subsection{Wrapped fluxed D6 branes}

In this section we analyze the stability of probe fluxed D6 branes on black hole backgrounds which have either purely electric or purely magnetic charges. We performed a KK reduction from 11d to 10d, the details of which are in Appendix \ref{sec:IIA}.

Before beginning with the probe analysis, a few comments are in order. First of all let us mention that the M-theory compactification we are using leads to a RR flux in IIA proportional to $J$, with $k=1$. This means that the 11d supergravity approximation is well-suited, however we will use perturbative IIA nomenclature “D6, D4, D2, D0" to label different charges in the 4D theory, even though we are not in a regime in which there is a reliable perturbative string description of the corresponding sources as D-branes \footnote{Even in vacua in which IIA string perturbation theory is valid, objects with D6-D4-D2-D0 charges large enough to be well-described as weakly-curved black hole geometries in supergravity necessarily fall outside the regime of applicability of string perturbation theory. }. In the present setup, it would perhaps be more apt to refer to these charges as “KK-monopole, M5, M2, KK mode", but we will stick to the former nomenclature. We will moreover assume that the computation of the masses by using the IIA Born-Infeld action, imagining the “D6, D4, D2, D0" charges as arising from wrapped D-branes, is valid. Along this reasoning, at large charges and large $l_{AdS}$ one is bound to retrieve the asymptotically flat case, and we find that this is the case.  

Another important point is that we want the fluxed D6 branes to be treatable as “probes". A fluxed D6 will be very heavy at $k=1$: $m_{D6} \sim N^{3/2} / l_{AdS}$.%
\footnote{%
	The compactification on $M^{111}$ is just one example in a class of compactifications on manifolds called $M^{pqr}$, labeled by 3 nonnegative integers \cite{Witten:1981me}.
	These are not all independent, since $M^{pqr} = M^{pq0} / \mathbb{Z}_l$ where $l = (3 p^2 + q^2) / k$ where $k$ is is the highest common factor of $2rp$, $rq$ and $3p^2 + q^2$ \cite{Castellani:1983yg}.
	Furthermore, when $p = q$ and $r \ne 0$, the low-energy effective action of the resulting M-theory compactifications is independent of the values of $p$ and $r$ \cite{Fabbri:1999mk}.
	Therefore, taking $p=q$ and $r=1$, we have that $k = p$ and $l = 4p$, so we find that $M^{pp1} = M^{110} / \mathbb{Z}_{4p}$ and the volume of the compact manifold can be made arbitrarily small by taking $p$ large. 
}
For this to be treatable as a probe, it needs to be parametrically lighter still than the mass of the background black hole, which is of the order $M \sim r_H / G$ where $r_H$ is the horizon radius\footnote{We thank F. Denef for correspondence on these issues.}. In our setup, in which we kept fixed $l_{AdS} =1/2$ we were able to find stable probes on background black holes having $r_H =1 $, but 
for larger black holes, the stable probes appear only for increasingly low temperature, making it numerically more challenging. Intuitively, this effect is due to the fact that a larger black hole fills up the spacetime and the probe is pushed closer towards the horizon by the confining potential in AdS.
We also considered planar configurations (which can be viewed as the large-size limit of spherical black hole configurations) and found solutions with metastable probe D6 branes outside the horizon.

Last but not least, let us mention that there are subtleties regarding the allowed boundary conditions for baryonic operators in ABJM-like theories, see for example \cite{Klebanov:2010tj}, with further progress in the recent paper \cite{Bergman:2020ifi}. We will come back to both these points later on in the discussion section.

\vspace{3mm}

Being aware of the aforementioned subtleties, and of the regime of validity of our computations, we can proceed with the calculation of the DBI action, which we report in App. \ref{sec:IIA}. It turns out that it assumes the familiar form
\be \label{probe_final}
S_{D6,f} = \int dt \,\, m_{\Gamma} (t^i) \sqrt{g_{tt}} + \frac{1}{2\sqrt2} \int (\mathbf{Q}_I A^I - \mathbf{P}^I B_I)  \,, \qquad m_{\Gamma}(t^i) = |\mathcal{Z}(t^i, \Gamma)| \,,
\ee
where $\Gamma$ is the probe charge, which is a function of the worldvolume flux $\mathbf{f}$, $\mathcal{Z}(t^i, \Gamma)$ is the central charge of the ungauged theory, and $A^I$ and $B_I$ are the electric and dual magnetic potentials respectively (see the definition in Appendix A).

First of all, one can notice that the hypermultiplets have dropped out from the expression of the mass, which depends solely on the vector multiplets scalars. Previous examples of probe computations (wrapped M2 branes on AdS$_4$ backgrounds of \cite{Klebanov:2010tj}, and wrapped D3 branes on AdS$_5$ black branes discussed in \cite{Henriksson:2019ifu}) also showed the same pattern. It is however surprising that this sector of the theory does not contribute to the mass term of the DBI action.

The expression \eqref{probe_final} coincides with the probe action for a black hole in ungauged $\mathcal{N}= 2$ supergravity, where BPS extremal black holes have mass $ m =|\mathcal{Z}|$. In \cite{Billo:1999ip} it was shown that in Calabi--Yau compactifications, the DBI action for a D-brane wrapped on supersymmetric cycles in the internal manifold gives exactly \eqref{probe_final}, via kappa-symmetry, justifying its use as effective action for a probe black hole in asymptotically Minkowski spacetimes. A probe action of the form \eqref{probe_final} was postulated in \cite{Anninos:2013mfa} and was rederived via the DBI action in \cite{Asplund:2015yda} for the M-theory compactification on $S^7/ \mathbb{Z}_k$. Let us finally mention that an expression of the form \eqref{probe_final} is compatible with our intuition since the  mass  of  an  extremal  black  hole  becomes insensitive  to  the  AdS curvature for sufficiently large AdS radius. 

In what follows we will use the explicit value of the central charge
\begin{eqnarray}
|\mathcal{Z}(t^i, \Gamma)| & = &  | \textbf{f} ^3-i \textbf{f} ^2 \tau_1-\textbf{f}  \tau_1 \tau_2+i \tau_1 \tau_2 \tau_3-\textbf{f}   \tau_1 \tau_3-i \textbf{f} ^2 \tau_2 -\textbf{f}  \tau_2 \tau_3-i \textbf{f} ^2 \tau_3 | \,, \end{eqnarray}
and all the computations in this section will take place in the rotated symplectic frame called “magnetic $STU$" ($mSTU$)
\be
F_{mSTU} = -2i \sqrt{X^0 X^1 X^2 X^3} \,.
\ee
As explained in Sec. \ref{4dSuGra}, this allows us to avoid working with tensor multiplets. The charges and the sections in the two frames are related via the symplectic rotation \eqref{ss_rotation},
therefore, the probe charge for a pure fluxed D6 in the model with $F_{mSTU}$ is
\be \label{gammarotated}
\Gamma = (P^0, P^1,P^2,P^3,Q_0,Q_1,Q_2,Q_3) = \left( 1, -\textbf{f} ^2, -\textbf{f} ^2 , - \textbf{f} ^2 , -\textbf{f} ^3 , \textbf{f} , \textbf{f} , \textbf{f}  \right)\,.
\ee

\subsubsection*{Strings attached to the D-branes \label{stringsattached}}

In the reduction to IIA under consideration, D2 and D6 branes come with fundamental strings attached due to the fact that they have a magnetic charge with respect to the massive $U(1)$ vector field \cite{Aharony:2008ug}.

One can see this from the equation for the worldvolume gauge field \eqref{eq:WV} (see also the discussion in Section 4.4 of \cite{Asplund:2015yda})
\be
d \star_7 (F + \ldots) = dC_5 + L_s^2 dC_3 \wedge F + \frac{L_s^4}{2} dC_1 \wedge F \wedge F \,,
\ee
on the D6 brane worldvolume.
The integral of the left hand side vanishes and so must the right hand side.
Using the the expressions for $C_1$, $C_3$ and $C_5$ in Appendix \ref{sec:IIA} (\ref{Aa}, \ref{eq:dC3}, \ref{eq:dC5}), one can see that the contribution from $dC_3$ has vanishing pullback onto the worldvolume, whereas from $dC_5$ only the contribution $\star_{10} H_4$ survives. Integrating both sides, using the flux \eqref{worldvol}, this gives
\be
0 =  -6 \times \frac{1}{8^3} \times (4 \pi)^3 + \frac62  \times \frac14 \times \left( \frac18 \right)^2 \times (4\pi)^3 \textbf{f}^2 =  -6 \frac{(\pi)^3}{8} +  \frac34 \pi^3 \textbf{f}^2
\ee
which is satisfied if $\textbf{f}^2= \pm1$.

One can see this from the four-dimensional point of view, by noticing that the massive vector combination is\footnote{Notice that we changed frame, dualizing the two form into the pseudoscalar $a$, and effectively switching the prepotential $F = 2 i\sqrt{X^0 X^1 X^2 X^3}$.}
\be
\mathcal{B} = 6 A^0 +2 A^1 +2 A^2 +2A^3 \,.
\ee 
Computing its magnetic field strength and integrating to get the charges we obtain 
\be
0 = 6P^0 + 2 P^1 + 2 P^2 +2P^3 =  6 -3 \times 2 \textbf{f}^2 = 6 (1- \textbf{f}^2) \,,
\ee
where in the second equality we used \eqref{gammarotated}. This means that for the value of worldvolume flux $\textbf{f}^2 =1$, the brane configuration is consistent and the tadpole is cancelled.  We will see in what follows that the value $\textbf{f}=1$ gives stable probes, and we stick to this value of worldvolume flux in most of our computations, knowing that for a different value we would need to take into account the strings stretched between the probes (nevertheless, we plot different values of $\textbf{f}$ in Fig. \ref{FigDouble} to analyze the dependence of the potentials on the probe charges).

We will now proceed to plot the behaviour of the effective potential for various black hole backgrounds, focusing first on spherical purely electric solutions (with massive vector halo), and on magnetic black holes with hyperbolic horizons which admit a supersymmetric limit. For the latter configurations, the massive vector field is switched off: this is due to the spherically symmetric ansatz that we adopt. To anticipate the results in the next two sections, we will see that  there exist stable and metastable pure fluxed D6 probes at finite (small) temperature on the electric backgrounds, while we do not find such stable probes on the magnetic black holes.

\subsubsection{Purely electric background black hole \label{sec_D6probes}}

\begin{figure}[H]
\begin{center}
    \includegraphics[width=\textwidth]{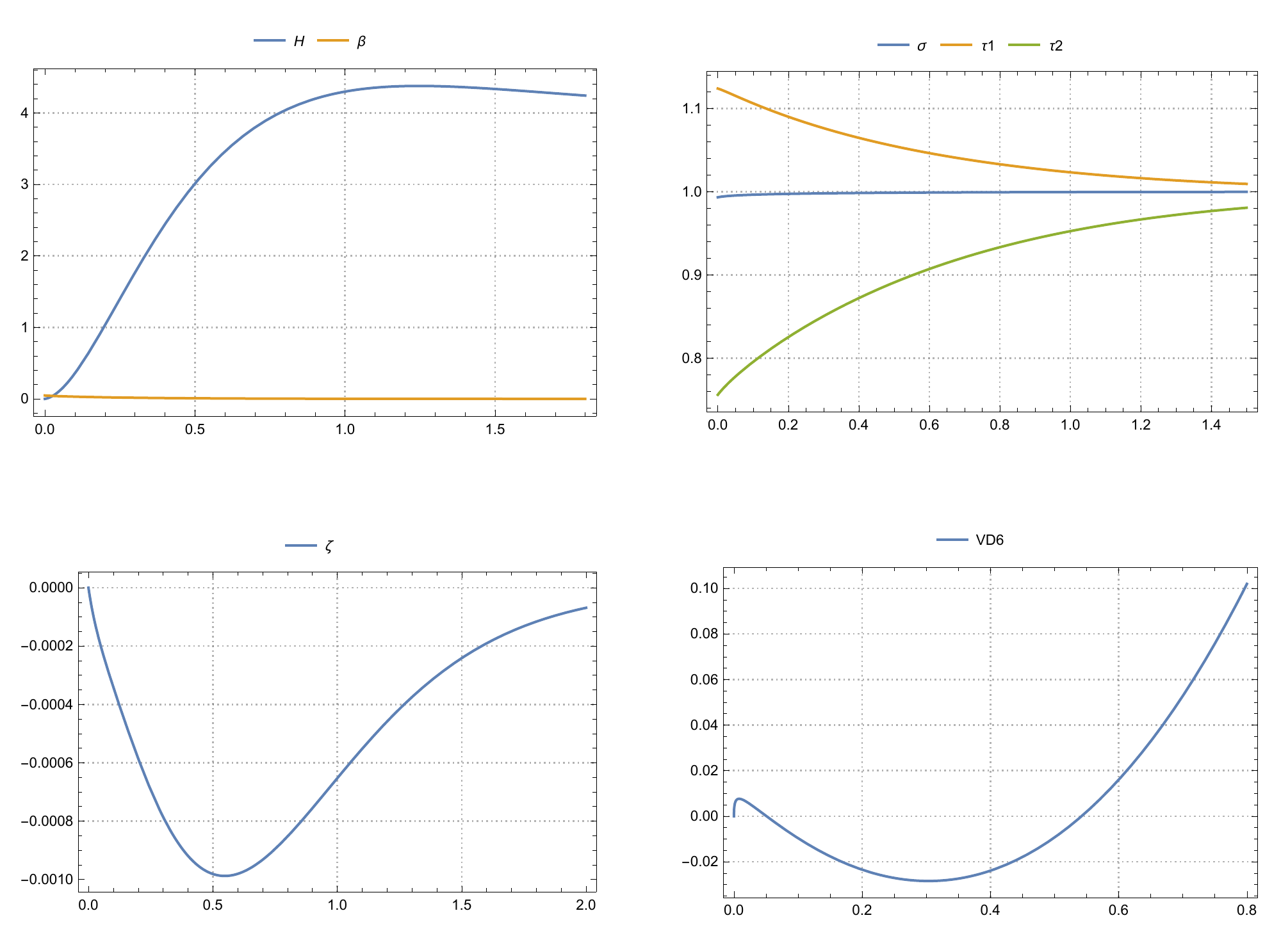}
    \end{center}
    \caption{Radial profile of the fields for a purely electric solution with parameters $p\xi_1= p\xi_3 = -0.66$, $ p\xi_2 = -0.089$
    and $r_H=1/4$, with worldvolume flux $ \textbf{f}  =1 $ for the probe D6 brane. 
    This solution has temperature $T = 0.013$. The potential $V_{D6}$ for fluxed D6 branes dips below zero, signalling the presence of stable probes. \label{Fig_profile_potential}}
  \end{figure}

\begin{figure}[H]
\begin{center}
    \includegraphics[width=150mm]{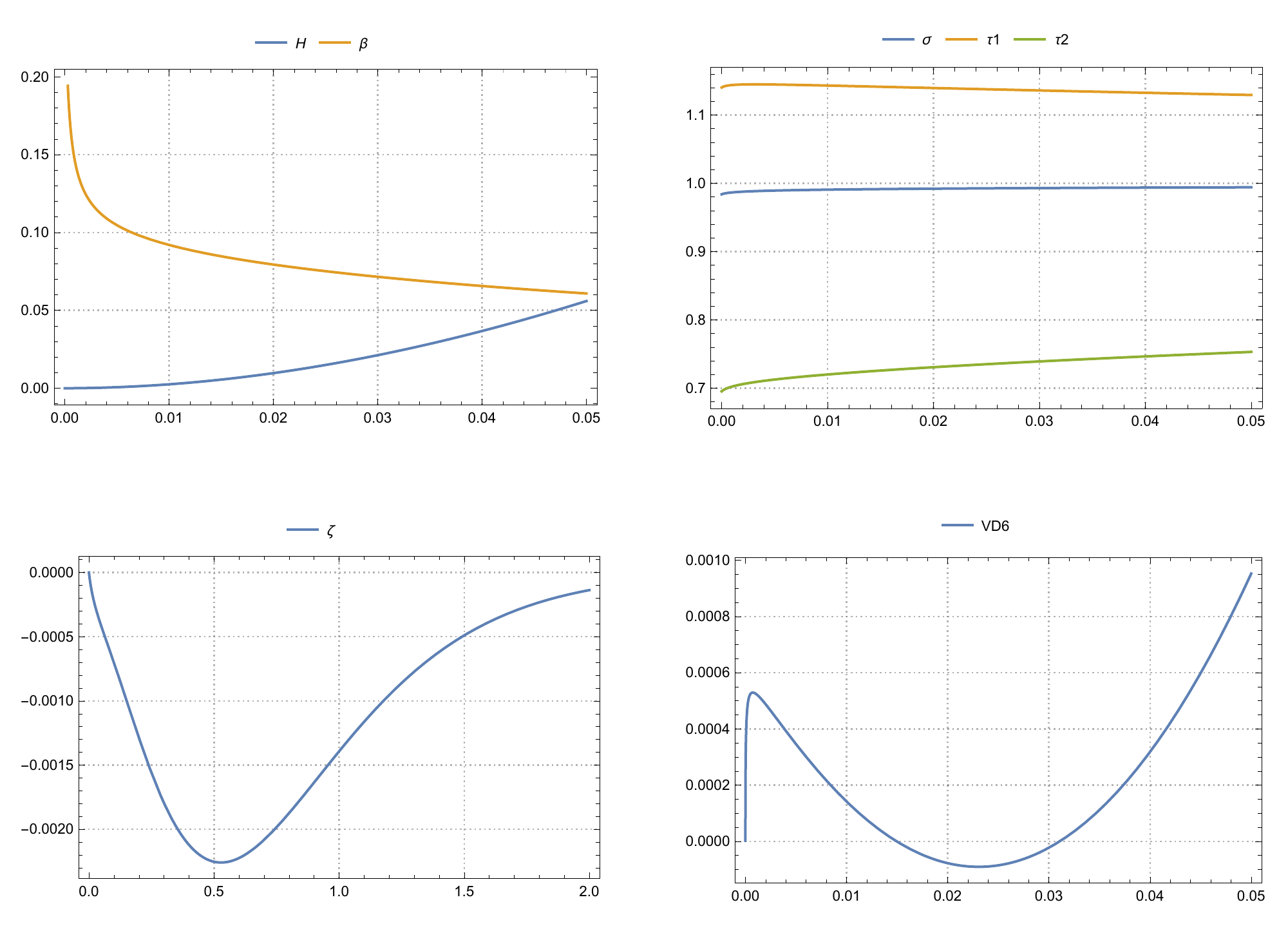}
    \end{center}
    \caption{Radial profile for the electric solution with parameters $p\xi_1= p\xi_3 = -7.35$, $ p\xi_2 = -0.99$    and $r_H=1$, with worldvolume flux $ \textbf{f}  =1 $ (notice the different scale of the horizontal axis for the profile of the massive vector field). The temperature of this black hole is $ T = 0.000476$.
    \label{Fig_profile_potential1}}
  \end{figure}

In this section we consider purely electric background solutions, characterized by a nonvanishing massive vector field halo. To find these numerically, we start with the Reissner--Nordström at given $r_H$ and $Q_1$ and increase the other charges step by step. We thus find a series of solutions with decreasing temperature for which the fluxed D6 brane potential $V_{D6}$ develops a minimum outside the horizon, first metastable and then stable. For black holes with larger radii, it is necessary to go closer to extremality for the minima of the potential to appear. We show solutions with $r_H= 1/4$ and $ r_H=1$ in Figures \ref{Fig_profile_potential} and \ref{Fig_profile_potential1}.
In Fig. \ref{FigDouble} we compare cases with different temperature and worldvolume flux.

  
We have not done a complete scan of the 6-dimensional parameter space of spherical electric black holes, see \Cref{tab:dofCounting}. In particular we have kept the leading modes of light scalars turned off and $Q_3 = Q_1$. By sampling this
space of solutions we were able to draw the following conclusions:

\begin{itemize}
\item Consistently with the expectations, a sufficient increase in the temperature makes the stable probes disappear (see Fig. \ref{FigDouble}). 
Notice that in order to find bound states we were inspired by the existence of stable probes, namely we have focused our search to the equivalent of the green region of Fig. 5.2 of \cite{Anninos:2013mfa}\footnote{A direct mapping between the model we use and that implemented in \cite{Anninos:2013mfa} is not possible since, among other things, the number of vector multiplets is different. We have however reproduced their analysis in the STU model in Appendix \ref{sec:S7comparison}: we have set one of the vectors to zero and taken this as a starting region where to possibly find stable probes.}. 
\item We have found that there is a specific range of values for the worldvolume parameter $\textbf{f}  \in [\textbf{f} _{a},\textbf{f} _{b}]$ that allow presence of stable and metastable probes. For instance, for the solution in Fig. \ref{Fig_profile_potential}, the interval in which the probes are stable is $\textbf{f}  \in [0.644, 1.648]$, while metastable are in $\textbf{f}  \in [0.561, 0.644[$ and in $\textbf{f}  \in ]1.648, 2.04]$. Outside this region, the probes become unstable. Notice that the value $\textbf{f} =1$, which gives zero strings attached to the probes, lies in the interval which allows for stable probes.

 \begin{figure}[H]
\begin{center}
    \includegraphics[width=73mm]{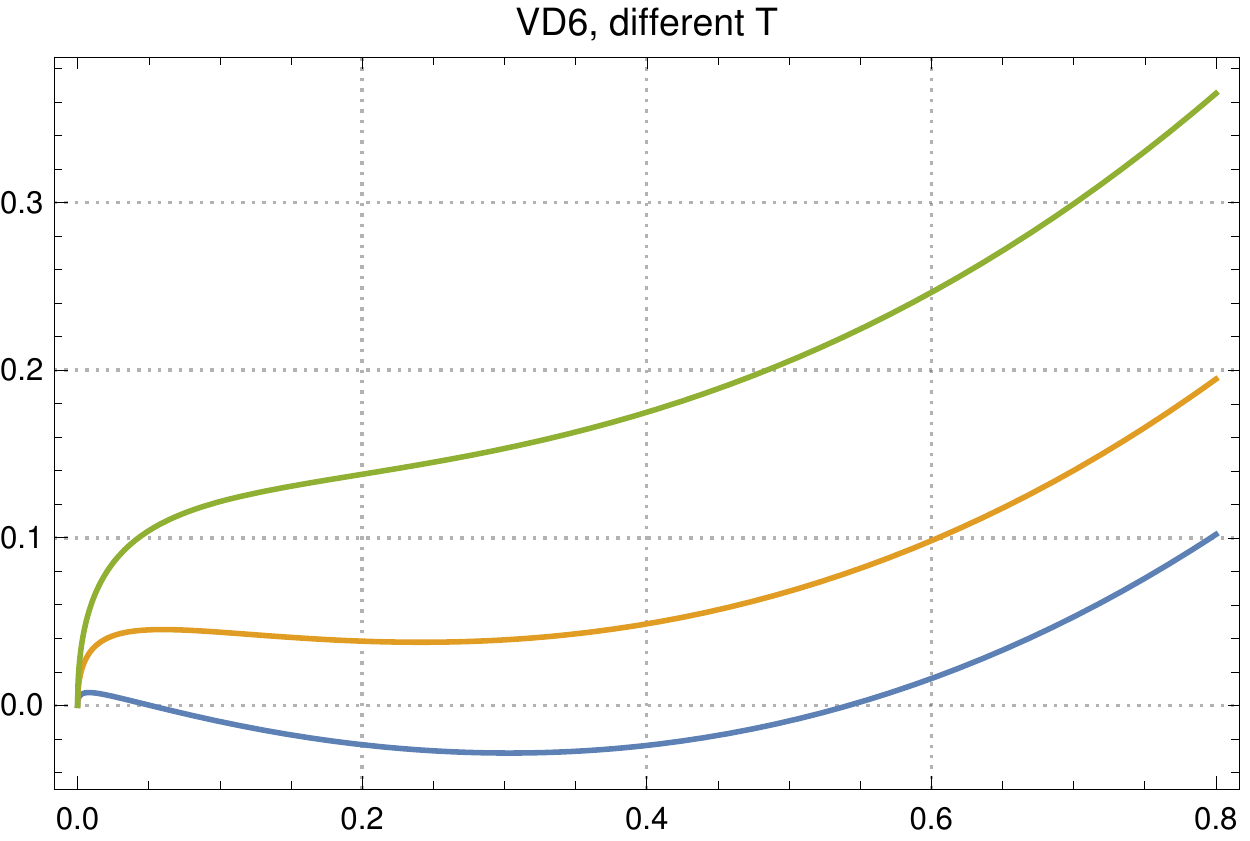}
    \includegraphics[width=73mm]{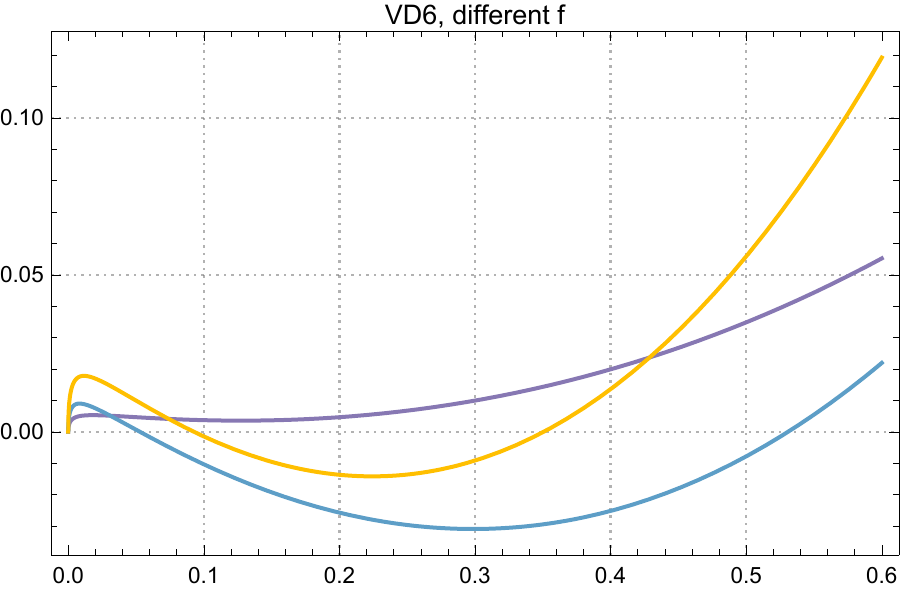}
    \end{center}
    \caption{D6 probe potential for probes with worldvolume flux $ \textbf{f}  =1 $  on black holes  with $p\xi_1= p\xi_3 = -0.66$, $ p\xi_2 = -0.089$, and different temperatures:  $T = 0.013$   (blue, $r_H=1/4$), $T = 0.07$ (orange, $r_H=26/100$), $T = 0.16$ (green, $r_H=28/100$). On the right, solutions with $ T = 0.013$, and different values $\textbf{f} =0.6$ (violet), $\textbf{f} =1.1$ (light blue) and $\textbf{f} =1.5$ (yellow).\label{FigDouble}}
  \end{figure}
  
\item Increasing the value of the parameter $r_H$ in the black hole background makes stable and metastable probes lie closer to the horizon. This is also reflected in the two plots in Fig. \ref{Fig_profile_potential} and \ref{Fig_profile_potential1} (notice the different scales on the horizontal axes). It is due to the fact that a bigger black hole “fills up" AdS space, and the confining potential of the negative cosmological constant pushes the probes closer to the center. Since the limit of large spherical black holes should be captured by planar solutions, we have considered those as well. As shown in \Cref{fig:planarElectricD6} we found metastable probes in these configurations for non-integer values of the flux $\mathbf{f}$. We were unable to find solutions with fully stable probes as well as metastable probes with $\mathbf{f}=1$, but it is possible that such solutions exist at lower temperatures.

\item In the cases we examined, the radial location of the minimum of the potential is located further away from the horizon for stable probes than for metastable ones, which instead seem to be pushed closer to the horizon - see for instance Fig. \ref{FigDouble} and Fig. \ref{FigHalo}. In “large" background solutions ($r_H > l_{AdS}$) the stable probes lie in the region within the horizon and the “peak" of the amplitude for the massive vector halo, while for smaller black holes these lie further away, on top of the peak. Generically, the place where the stable probes reside roughly coincides with the region where the massive vector field is more pronounced. It would be interesting to check if this is the case also for other kind of probes or massive objects (e.g.~probe black rings).

\item Metastable probes which lie closer to the horizon have lower “barrier heights" with respect to  the ones lying further away. This seems to be in agreement with what was found in \cite{Anninos:2013mfa}. We make further comments on this point and on the relaxation dynamics of metastable probe clouds in the conclusions.

\end{itemize}

\begin{figure}[H]
	\begin{center}
		\includegraphics[width=149mm]{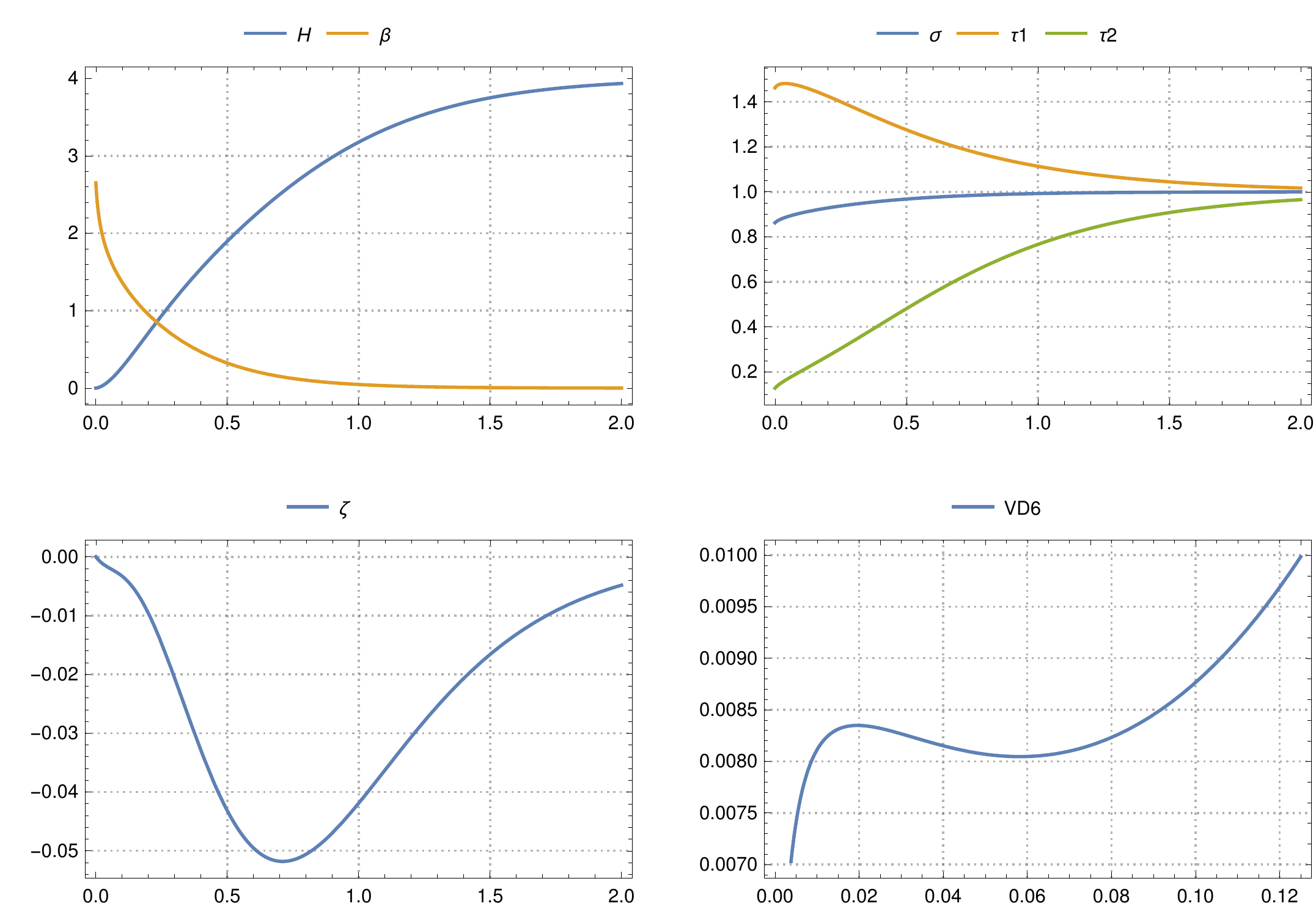}
		\caption{Planar solution with purely electric charges $p\xi1 = -5.58$, $p\xi2 = -2.84$ and $r_H = 1$. This solution has temperature $T = 0.01$. The flux through the D6 brane probe is given by $\mathbf{f} = 0.66$. Probe D6 branes with $\mathbf{f}=1$ are unstable on this background solution. Notice the different scale of the horizontal axis for the bottom right figure.}
		\label{fig:planarElectricD6}
	\end{center}
\end{figure}

 \begin{figure}[H]
\begin{center}
        \includegraphics[width=75mm]{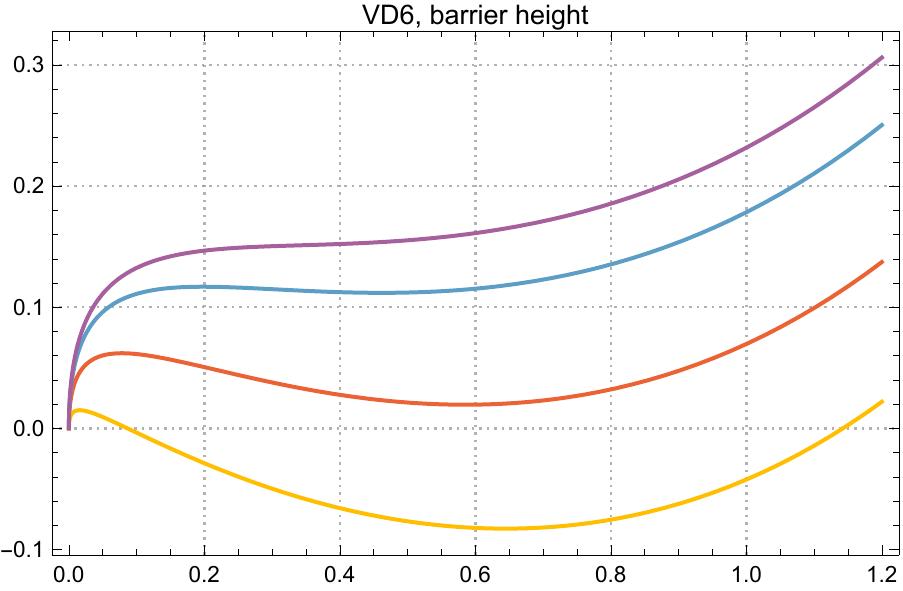}
    \end{center}
    \caption{Plot of the potential profiles for various probes on purely electric black holes with $r_H=1/8$, $\mathbf{f}=1$ and temperature $T=0.045, 0.17, 0.28, 0.33$ (yellow, red, blue, violet respectively). One can see that the probes closer to the horizon have a higher value of the minima of the potential (“barrier height") with respect to those lying further away. \label{barrier_heights}}
  \end{figure}

 \begin{figure}[H]
\begin{center}
    \includegraphics[width=77mm]{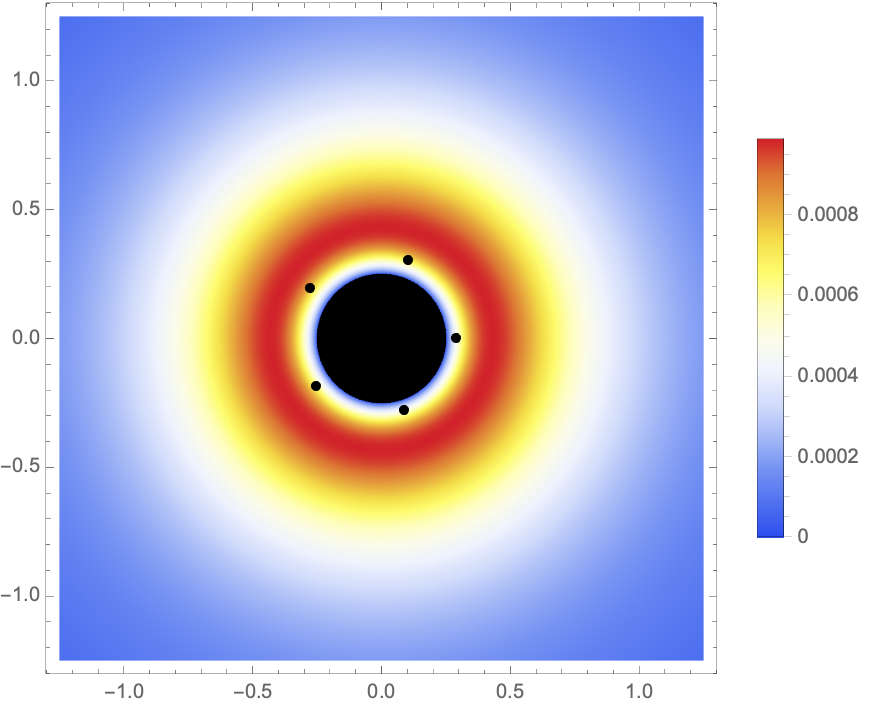}
        \includegraphics[width=77mm]{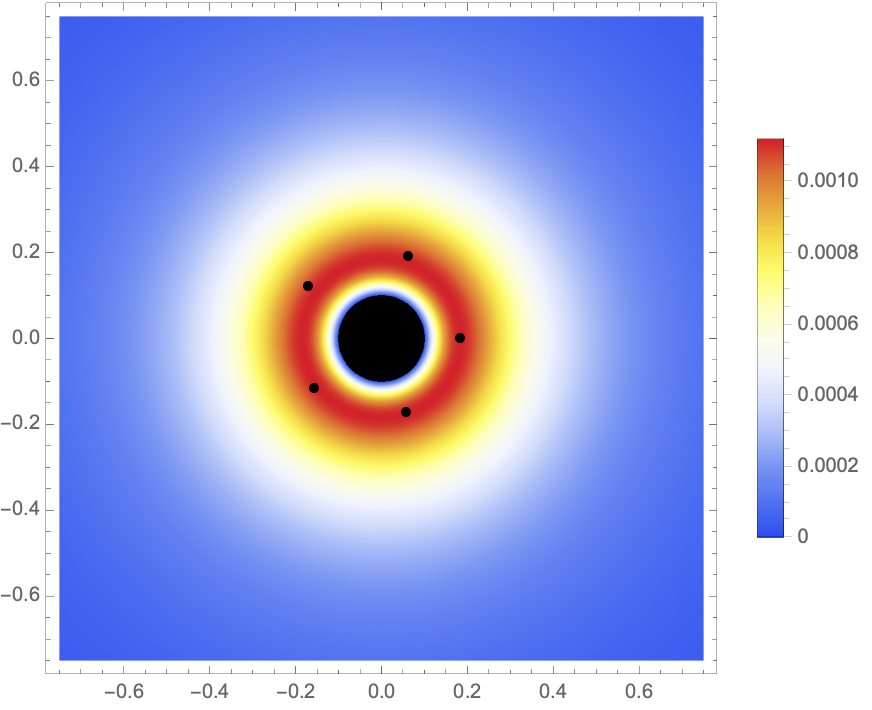}
    \end{center}
    \caption{On the left, the 2D plot of the massive vector field profile for the electric solution of Fig \ref{Fig_profile_potential}. The metastable probes (with $\textbf{f} =5/8$, rightmost, and $\textbf{f} =9/5$ the lowest one) lie close to the horizon, between the peak of the massive vector and the horizon. The other probes, starting from $\textbf{f} = 3/4$ (upper), $\textbf{f} =1$ leftmost,  $\textbf{f} =3/2$ bottom left are all stable, and again their equilibrium distances lie behind the massive halo. Right: 2 metastable (rightmost, lowest) and 3 stable probes on the background solution characterized by $p\xi_1= p\xi_3 = -0.194$, $ p\xi_2 = -0.026,$ $T = 0.15$  and $r_H=1/10$. We see that in this case the equilibrium distance for stable probes lies on top of the peak of the massive vector. Notice that only the probes with $\textbf{f}=1$ come without strings attached. \label{FigHalo}}
  \end{figure}

\subsubsection{Asymptotically flat limit and “caged wall crossing"}

We can consider the small black hole/asymptotically flat limit by sending the cosmological constant to zero, i.e. taking the gauge coupling constant to zero. This corresponds to switching off the mass term for the vector field as well, therefore the analysis follows that of \cite{Anninos:2013mfa}. We recap it here for simplicity, before analyzing the case of probes on 1/4 BPS black hole backgrounds. 

\begin{figure}[H]
\begin{center}
    \includegraphics[width=75mm]{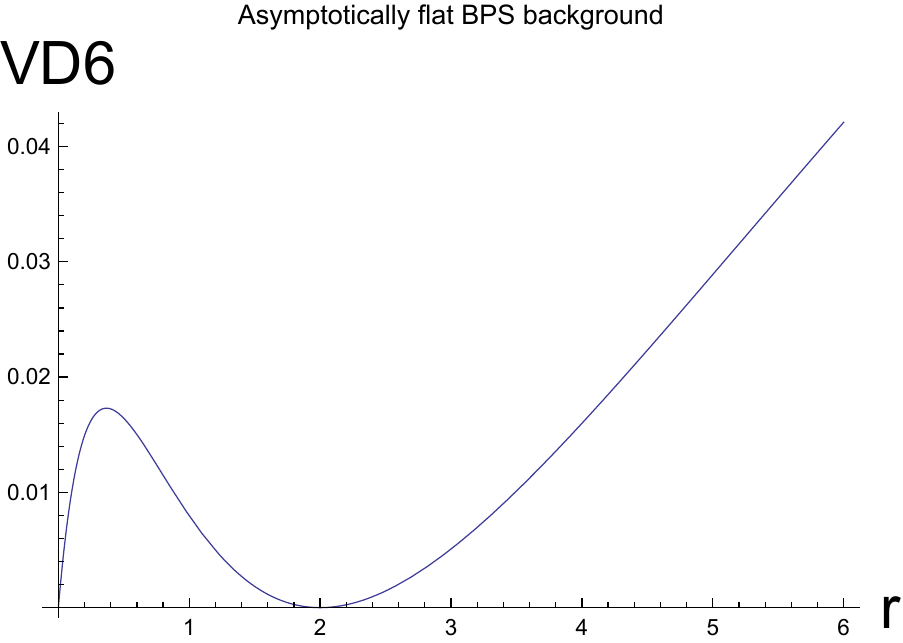}
    \end{center}
    \caption{BPS potential for D6 probes in asymptotically flat space: there is a minimum of the potential, $V=0$ outside the horizon at the value $\text{x}_{BPS} =2$ (charges are $q_0 = 7 , q_1 =q_2 =q_3= 1$, worldvolume flux $\mathbf{f} =1$ and the asymptotic values of the scalars are set to one). The value of $\text{x}_{BPS}$ obtained from the probe potential agrees with the value of the exact two center solutions of \cite{Bates:2003vx}. \label{potential_minkowski}}
  \end{figure}

Sending the gauging to zero corresponds to the framework of ungauged $\mathcal{N} =2$ supergravity, where multicenter BPS black holes preserving half of the supersymmetries are well known and studied \cite{Denef:2000nb,Bates:2003vx}. Analytic expressions are available for these configurations, and in the two-center cases one can easily find the radius of separation between the two centers (“BPS distance") of charges $\Gamma_1$ and $\Gamma_2$, according to this formula:
\be
\text{x}_{BPS} = \frac{\langle \Gamma_1 ,\Gamma_2 \rangle}{ 2 \text{Im} (e^{-i \alpha} \mathcal{Z}(\Gamma_1))_{\infty}} \,,
\ee
where $\mathcal{Z} =  \langle \mathcal{V} , \Gamma \rangle $ is the central charge of the $\mathcal{N} =2$ theory. More explicitly, in the symplectic frame that we have chosen (which is rotated with respect to that used in \cite{Anninos:2011vn}), for a solution with $Q_1=Q_2=Q_3$ the formula reduces to
\be \label{xBPS}
\text{x}_{BPS} = \frac{ Q_0- 3 \mathbf{f}^2 Q_1}{3a_0 \mathbf{f}^2-a_1} \,,
\ee
where $a_0$ and $a_1$ are the asymptotic values of the symplectic sections $X^0$ and $X^1$ (see notation of \cite{Anninos:2011vn}, which we follow with different normalizations). The BPS equilibrium distance can be read off from the  potential $V_p$ for a fluxed probe D6 brane on background of an asymptotically flat BPS black hole, which is of the form (see for instance \cite{Anninos:2011vn}) 
\be V_p = \sqrt{\Delta^2 + V_{em}^2} + V_{em} \,.
\ee 
Let us focus on a supersymmetric probe black hole. For some radial coordinate there is a minimum of the potential at $V_p =0$: this is due to the fact that $\Delta_{BPS} =0 $ and $V_{em} <0$. The condition $\Delta_{BPS} =0$ gives exactly formula \eqref{xBPS}. The situation is depicted in Figure \ref{potential_minkowski}, where one can see that the potential has a zero for $r = \text{x}_{BPS}$.
Notice that the quantity $\text{x}_{BPS} $ diverges for $ \frac{3a_0 \textbf{f}^2}{a_1} =1 $. This location in moduli space corresponds to decay at marginal stability: the distance between centers becomes infinite and the bound state ceases to exist. 

Anti-de Sitter space acts like a box and has a “confining" effect: in this setup, the true separation distance cannot diverge and for a sufficiently large radius the bound state will start feeling the presence of the AdS potential\footnote{In addition to this, in the supergravity models in which BPS solutions have been found the value of the scalar fields is fixed asymptotically demanding that it extremizes the scalar potential. Therefore there are no moduli to tune that would allow the phenomenon of wall crossing.}. In \cite{Anninos:2013mfa} an expression for the first correction to the potential due to the AdS confining force was given. The solutions treated in that case are analytic and in a closed form, which makes the analysis more transparent. The upshot of their finding is that when we vary the parameters such that we pass through the location where $\text{x}_{BPS}$ diverges, no actual decay will happen. Instead, the minimum of the potential is lifted to give a metastable probe. 

In our case, 
we have verified numerically that the probe potential has the same qualitative behaviour: the value\footnote{In our case the value of the scalar fields at infinity are fixed to one. The value of the worldvolume flux $\mathbf{f}= 1/\sqrt3$ which gives a diverging $\text{x}_{BPS}$ lies in the range of metastable probes (see previous section). Notice however that, as mentioned in Sec. \ref{stringsattached}, for this value of worldvolume flux the probes has fundamental strings attached to it.} of worldvolume flux for which $\text{x}_{BPS} \rightarrow \infty$ leads to a metastable probes. 
Note that even though we discussed what happens upon taking the ungauged supergravity limit $g \rightarrow 0$ and expanding around the 1/2 BPS solution, there are actually no known regular static solutions preserving half of the supersymmetries in the gauged supergravity models under consideration: all static known solutions of this kind (i.e. \cite{Duff:1999gh}) present naked singularities. However there exist static configurations of finite area that preserve 1/4 of the supercharges. In the next section we use the latter as backgrounds for the probe analysis.

\subsubsection{Probes on purely magnetic supersymmetric  black hole background \label{susy-magn}}

In this section we are interested in investigating the stability of D6 probe branes on top of backgrounds that have a well-behaved supersymmetric limit, corresponding to the Halmagyi-Petrini-Zaffaroni  BPS black holes \cite{Halmagyi:2013sla}. The details of these solutions can be found in Appendix \ref{susyHPZ} and here we recap their salient features, focussing in particular on purely magnetic configurations with no axions. 

\begin{figure}[H]
\begin{center}
    \includegraphics[width=75mm]{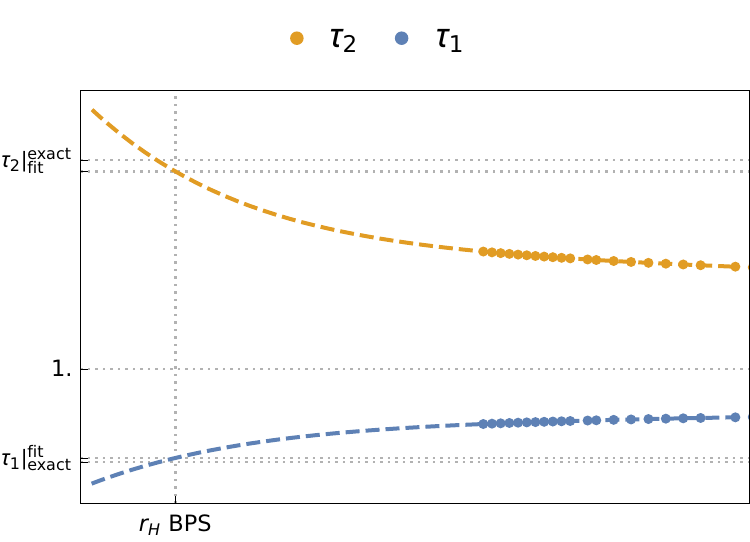}
    \includegraphics[width=75mm]{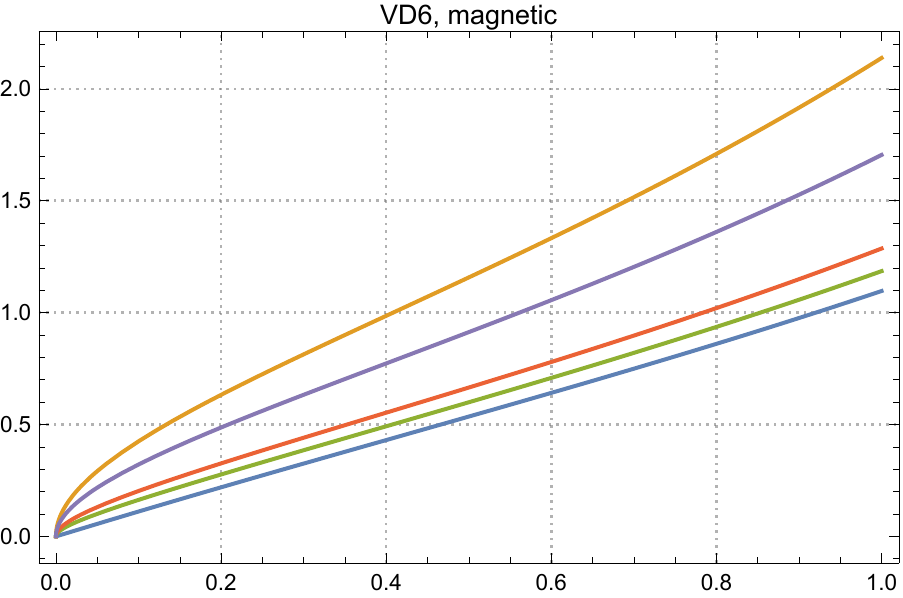}
    \end{center}
    \caption{On the left, extrapolated values for $\tau_1$ and $\tau_2$. Note that the horizontal axis is not the $u$ coordinate within a single solution, but rather the value of $r_H$ that labels different solutions: each of the solid points indicates a numerical solution. The lowest temperature we obtain is $T \sim 0.0003$. On the right: Probe potential for solutions with magnetic charge that approximate the BPS solution of \cite{Halmagyi:2013sla}. The temperatures are: $T= 2.55$ (orange), $T= 1.71$ (violet), $T= 0.69$ (red), $T = 0.16$ (green), $T= 0.00038$ (blue). From our analysis we see no sign of stable or metastable probes.  \label{magnetic_noinfl}}
  \end{figure}

Supersymmetry, coupled with the requirement of having a finite nonzero entropy, forces us to work with a hyperbolic horizon. These solutions have a zero-temperature, supersymmetric limit characterized by a near-horizon geometry of the form $AdS_2 \times \mathbb{H}^2$. The explicit near-horizon geometries (and a numerical solution interpolating between the near-horizon geometry and an AdS$_4$ asymptotic region) were first found in \cite{Halmagyi:2013sla}. The near-horizon metric reads
\be
ds^2 =  - R_1^2 r^2 dt^2 + \frac{dr^2}{ r^2 R_1^2} + R_2^2 (d\theta^2 + f(\theta)^2 d\phi^2) \,, \qquad \quad f(\theta) = \left \{ \begin{array}{cc} 
\sin \theta \,\,\, \text{for} \,\,\, \kappa =1 \\
\sinh \theta \,\,\, \text{for} \,\,\, \kappa =-1 \end{array} \right.
\ee
with gauge fields
 \be
 A^{\Lambda}=  \tilde{q}^{\Lambda} r \,dt + P^{\Lambda} f'(\theta) d \phi \,,
 \ee
 \be
 \tilde{q}^{\Lambda} = -\frac{1}{R_2^2} \left(   \im \cN^{\Lambda \Sigma}  \re \cN_{\Sigma \Gamma} P^{\Gamma}+ \im \cN^{\Lambda \Sigma} Q_{\Sigma} \right) \,.
 \ee
 We focus on solutions with zero axions $b_i=0$ and purely magnetic charges, $Q_{\Lambda} =0$. The magnetic charges are
\be
 P^0 = -\frac{1}{4 \sqrt2}\,, \qquad  P^3 = P^1\,, \qquad P^2 = \frac{3}{4\sqrt2} -2 P^1\,, \qquad P^1 =  -\frac{ \tau_1^2 \left( \tau_1^2-3\right)}{4 \sqrt{2} \left(\tau_1^2+1\right)} \,,
\ee
with
\be
 \tau_3 =\tau_1\,, \qquad  \tau_2 = \frac{3- \tau_1^2}{2 \tau_1} \,,
 \ee
where for simplicity we expressed the charges in function of the horizon value of the scalar $\tau_1$.  
The size of the $AdS_2$ and the $S^2$ are
\be
R_1^2 =\frac{\tau_1}{32}  \left( 3 - \tau_1^2\right)\,, \qquad R_2^2 = -\frac{\kappa \tau_1 \left(\tau_1^4-2 \tau_1^2+9\right)}{32 \left(\tau_1^2+1\right)} \,.
\ee
After verifying that the (supersymmetric) near horizon geometry satisfies our equations, we implemented our shooting technique and found a series of purely magnetic full-flow black hole solutions with decreasing temperature, the near horizon geometry of which eventually approximates to a good extent the BPS near horizon geometry of \cite{Halmagyi:2013sla}, see \Cref{magnetic_noinfl}. To give an idea of the precision to which we work, the extrapolated value of the scalars at the horizon are within $2\% $ of the actual BPS value computed analytically from the near horizon-geometry data. We retrieve the extremal black hole solution by solving the equations of motion, imposing the boundary conditions discussed in Sec. \ref{sec:asymptoticBehavior}. In order to get the $T=0$ supersymmetric solution it would be more efficient and appropriate to directly solve the first order BPS equations, and analyze carefully the boundary conditions dictated by supersymmetry, along the lines of \cite{Freedman:2013oja,Cabo-Bizet:2017xdr,Bobev:2020pjk}. Just like the BPS equations generate a subset of the solutions to the full equations of motion, the allowed boundary conditions would be particular a subset of those in \Cref{tab:dofOverview}. For the time being we proceed nonetheless with our shooting technique setting the leading fall-offs to zero, leaving generalizations for future work\footnote{Supersymmetry requires that the real and imaginary parts of the scalars are quantized differently (pseudoscalars are subject to regular boundary conditions, while the scalars obey alternative quantization). Moreover, it relates the fall-offs of different matter fields sitting in the same supermultiplet. We thank Chris Herzog and Nikolay Bobev for remarks on these points.}.

Once again, while a complete scan of the parameter space is almost impossible with our current techniques, we have plotted several examples of solutions with temperature as low as $T \sim 0.0003 $. In all cases we have considered we were unable to find stable or metastable probes: the potential did not show any sign of inflection - see Fig. \ref{magnetic_noinfl}. 

We performed a similar probe analysis or purely magnetic (spherical and hyperbolic) solutions from the $S^7$ reduction from M-theory, and the results are reported in Appendix \ref{sec:S7comparison}. Similarly, also in that case we did not find any stable probes. Stable probes in supersymmetric black hole backgrounds were found in \cite{Anninos:2011vn} in the context of asymptotically flat 1/2 BPS black holes in a box of finite volume. The absence in our case could be due to the different way supersymmetry is realized (a different projector in the Killing spinor allows to have a 1/4 BPS solution) and to the confining effect of the AdS potential.

To conclude this section, we remind the reader that, despite having considered (approximately) supersymmetric background solutions, the probe D6 branes most likely break supersymmetry. The type IIA background that we have obtained upon reduction of the 11-dimensional configuration on the coordinate $\psi$ (see Appendix \ref{sec:IIA}) breaks supersymmetry  \cite{Sorokin:1985ap,Petrini:2009ur}. For this reason we are certainly not in a position to claim that supersymmetric bound states do not exist. Moreover, our analysis considered only the simplified setup in which the background is without axions. Repeating this analysis including dyonic backgrounds can considerably alter the picture, but a complete treatment of this case is, at the moment, outside of the scope of this work.

\section{Conclusions}

In this paper we have analyzed the instabilities of black holes and black branes towards nucleation of various kinds of D-branes and M-branes. 

The main result is the existence of stable probe black holes composed by wrapping branes on noncontractible cycles in Sasaki--Einstein manifolds. We have analyzed in particular the case of D6 branes wrapped on the six-dimensional internal manifold obtained upon reduction along the M-theory direction, and we have shown that bound states exist for specific sets of charges and for sufficiently low temperature. It seems then that, with respect to \cite{Anninos:2013mfa}, the Higgsing of one of the vector fields does not hinder the existence of multicenter configurations in duals of ABJM-like theories. The location of the stable probe-black holes, with respect to the peak of the massive vector field, seems to lie either closer to the horizon (large black holes), or roughly coincide with it (for smaller ones). Another general feature is that metastable probes seem to lie closer to the horizon with respect to stable ones. Moreover, in our case the wrapped fluxed D6 branes have the special KK direction orthogonal to the M2 brane worldvolume direction. In the asymptotically flat case, in five dimensions, a similar nucleation instability corresponds to the BMPV black hole transforming itself into a black ring \cite{Bena:2011zw} (see also \cite{deBoer:2008fk,Elvang:2002br}). It would be interesting to make connection to the physical picture described in those papers.

Interestingly, the supersymmetric purely magnetic solutions in this compactification seem not to support stable probes. Let us also mention that the status of the microstate counting for supersymmetric configurations of these models is still work in progress, due to the fact that the baryonic charges seem to not be visible in the large $N$ limit and our configurations need to have these switched on. Recent advances on the topic can be found for instance in \cite{Azzurli:2017kxo,Hong:2019wyi,Hosseini:2019ddy}. We restricted our attention only to non-axionic solutions for simplicity, but it would be interesting to generalize the analysis to static dyonic, and stationary configurations, for example those found in \cite{Chong:2004na,Chow:2013gba,Gnecchi:2013mja} and their supersymmetric limits \cite{Hristov:2018spe,Hristov:2019mqp}. In presence of rotation (and genuinely complex scalar fields) we foresee a richer phase space for stable D brane probes\footnote{In the context of microstate counting, wrapped D3 branes in Euclidean signature have recently played a role in \cite{Aharony:2021zkr}. There it was shown that non-perturbative corrections due to the D3 branes rule out some of the extra gravitational solutions allowing for a precise one to one match with the field theory duals.}.

There are a few points that we have left for future investigation.
First of all, particularly important is the role of boundary conditions in these sort of supergravity truncations (see for instance discussion in \cite{Klebanov:2010tj}), which allow a specific set of M branes. Recent investigations \cite{Bergman:2020ifi} regarding the interplay between boundary conditions in gravity and allowed gauge invariant operators in the dual ABJM theory pointed out that various versions of the latter theory exist, characterized by different the gauge groups. One version in particular allows for gauge invariant di-baryonic operators, corresponding to D4 branes. It would be interesting to connect this story to our framework, which involves theories dual to the $Q^{111}$ truncation, and probe D6 branes. This point for sure deserves a more thorough investigation, which we leave for future study. 

Secondly, in our case we set the magnetic component of the massive vector field to zero, and this is a consequence of the static and spherically symmetric ansatz we adopted. We could alternatively decide to relax this spherical symmetry and work with axially symmetric solutions: this would allow for vortex-like configurations, in which magnetic flux lines are squeezed together by the Meissner effect. Examples of vortices emanating from rotating black holes have been found numerically for instance in \cite{Gregory:2014uca}. Finding stationary black hole backgrounds would entail solving coupled PDEs and would complicate our analysis considerably, but it would certainly be very interesting to see if the presence of this vortex solution, which microscopically corresponds to strings stretched between various branes \cite{Aharony:2008ug}, would alter the probe brane analysis. 

Third, for planar horizons we have found an instability towards nucleation of spacetime filling M2 branes. This is of the same form as in \cite{Henriksson:2019ifu}, where five-dimensional black branes in the conifold $T^{11}$ theory were found to be unstable towards the emission of spacetime filling D3 branes. In both cases this phase transition is triggered not by the baryonic charges, but by the R-symmetry. It would be good to investigate whether the Weak Gravity Conjecture, which states that the quantum corrections should decrease the relative mass to charge ratio, so that the extremal black branes can decay, plays any role in our analysis. Moreover, it would be interesting to examine for our backgrounds a decay channel known as brane-jet instability \cite{Bena:2020xxb}, along the lines of \cite{Suh:2021icf}.
 
Let us conclude with some observations. In \cite{Anninos:2013mfa} the tunnelling probability of metastable probes was related to the relaxation dynamics of glassy systems. In particular, the authors found that a cloud of (finely spaced) metastable probes exhibits logarithmic aging behavior, for which an observable $\cal{O}$ will evolve schematically like $ \mathcal{O}(t) - \mathcal{O}(t_0) = -\log(t/t_0)$, typical of glasses. In other words, there is no time translation invariance in the system, but rather a scale invariance. Their computation considered the time evolution of the probe number densities, and took into account that there is a correlation between distances from the horizon and barrier heights,  with  the  probes  closest  to  the  black hole having relatively low absorption barriers, and therefore short lifetimes, and those far away having the longest lifetimes. In our case, we find the same qualitative behaviour (see Sec. \ref{sec_D6probes}) regarding barrier height and horizon distance, and a more thorough analysis would be desirable to show the same logarithmic law. This, together with the computation of the viscosity via holography, from the cross section of low energy gravitons, as done in \cite{Policastro:2001yc}, would allow us to corroborate the picture that relates these multicentered bound systems to the physics of glass.

Lastly, in addition to the already mentioned applications to relaxation dynamics and glassy physics, we would also like to point out that a more careful study of bound states with supersymmetry would be desired, in relation to the phenomenon of wall-crossing for AdS$_4$ black holes. It would be necessary first of all to identify in this context what serves the role the moduli that control the decay at marginal stability, and a more thorough investigation of the supersymmetric indices (whose computation is problematic for single center black holes already) would be desirable.

 We hope to come back to all these points in the near future. 

\vspace{-4mm}

\section*{Acknowledgements}

The authors would like to thank Francesco Benini, Davide Cassani, Oscar Henriksson, Kiril Hristov, Hyojoong Kim, Silviu Pufu, Alessandro Tomasiello, Thomas Van Riet for useful discussions and correspondence. We thank Tarek Anous and Iosif Bena for useful comments on the draft of this paper. Finally, we are indebted to Frederik Denef for numerous discussions throughout the completion of this project, for elucidating some subtleties regarding the probe action and for giving feedback. CT acknowledges support from the NWO Physics/f grant n. 680-91-005. RM was supported by ERC Starting Grant 679278 Emergent-BH and NSF grant PHY-19-14412.

\begin{appendices}
\addtocontents{toc}{\protect\setcounter{tocdepth}{1}}

\section{Reduction to IIA and fluxed D6 probe action}
\label{sec:IIA}
\subsection{Reduction to IIA}
The reduction from 11d supergravity with metric $g_{\mu\nu}$ and three-form field $C_{\mu\nu\rho}$ ($\mu, \nu = 0,...10$) results in the type IIA supergravity theory  in ten dimensions with indices denoted by $\alpha,\beta = 0,...9$ whose bosonic fields are the metric, dilaton and Kalb--Ramond field, as well as the RR 1-form and 3-form
\be
g_{\alpha \beta}, \Phi, B_{\alpha \beta}, \qquad C_{\alpha}, C_{\alpha\beta\gamma}
\ee
The field $B_{\alpha \beta}$ is an antisymmetric two-form field, and the field $\Phi$ is a scalar (the dilaton, not to be confused with the hyperscalar $\phi$). 

We will formally reduce the Sasaki--Einstein manifold $Q^{111}$ along the coordinate $\psi$ to get a IIA supergravity theory. It is a well-known fact that the reduction along this coordinate does not lead to a supersymmetric background, see for instance \cite{Sorokin:1985ap,Petrini:2009ur}, but this will not be necessary for our purposes\footnote{Later on we are going to wrap a D6 brane on the six dimensional internal manifold of topology $S^2 \times S^2 \times S^2$. Like in the case analyzed in \cite{Asplund:2015yda}, the probe action will take a “BPS form" linear in the charges, even in absence of supersymmetry. Roughly speaking, this is possible because the geometry is large, so any possible corrections to the leading-order classical geometric brane energy formula, which happens to take the form $M=|Z|$, are small (one could say that this form is not so much “BPS" but rather “geometric" or “classical"). }. We use the standard decomposition \cite{Becker:2006dvp}
\be
ds_{11}^2 = e^{-2\Phi/3} g_{\alpha\beta} dx^{\alpha}dx^{\beta} + e^{4\Phi/3}(d \psi+ C_{\alpha}dx^{\alpha})^2 \,.
\ee
For the 11d truncation on $Q^{111}$ of \cite{Cassani:2012pj} we have: 
\be
ds_{11,C}^2 = \left( \sigma^2 \tau_1 \tau_2 \tau_3 \right)^{2/3}  \mathcal{K}^{-1} ds_4^2+ \left( \sigma^2 \tau_1 \tau_2 \tau_3 \right)^{-1/3} ds^2 (B_6) + \left( \sigma^2 \tau_1 \tau_2 \tau_3 \right)^{2/3} (\theta + A_0)^2 = \nonumber
\ee
\be
= \frac{\left( \sigma^2 \tau_1 \tau_2 \tau_3 \right)^{2/3}}{\tau_1 \tau_2 \tau_3} ds_4^2+ \left( \sigma^2 \tau_1 \tau_2 \tau_3 \right)^{-1/3} ds^2 (B_6) + \left( \sigma^2 \tau_1 \tau_2 \tau_3 \right)^{2/3} (\theta + A_0)^2 \,,
\ee
where we took into account that $ 8 \mathcal{K} =  e^{-K}$ and we remind the reader that 
\be
\theta = d\psi + \sigma_1 + \sigma_2 + \sigma_3 = d\psi + \frac14 ( \cos \theta_1 d\phi_1 +  \cos \theta_2 d\phi_2 +  \cos \theta_3 d\phi_3 ) \,.
\ee
The dilaton is then
\be
\Phi =  \frac12 \log (\sigma^2 \tau_1 \tau_2 \tau_3)\,.
\ee
From the 10-dimensional perspective, we have the metric
\be
ds_{10}^2 =  \sigma^2  ds_4^2+\frac{\tau_1}{8}  \left( d\theta_1^2 + \sin^2 \theta_1 d\phi_1^2 \right) + \frac{\tau_2}{8}  \left( d\theta_2^2 + \sin^2 \theta_2 d\phi_2^2 \right)+\frac{\tau_3}{8}  \left( d\theta_3^2 + \sin^2 \theta_3 d\phi_3^2 \right)\,,
\ee
and the Ramond-Ramond field $C_{\alpha}$, obtained from the metric component $g_{\mu \psi}$
\be \label{Aa}
C_{\alpha} = \sigma_1 + \sigma_2 + \sigma_3 + A_0\,,
\ee
where $A_0$ is the graviphoton.

Furthermore, we have the 10-dimensional three-form $C_{\alpha\beta\gamma}$, coming from the M-theory 3-form which can be decomposed into basis forms on the four-dimensional noncompact space
\be \label{CC3}
C_{\mu \nu\rho} = \mathcal{C}_3 + \tilde{B} \wedge (\theta + A^0) - A^i \wedge \omega_i + \xi^A \alpha_A -\tilde{\xi}^A \beta_A +b^i \omega_i \wedge (\theta +A^0)\,.
\ee
Here $\mathcal{C}_3$ is a three form (which in our case will only have components along the $AdS_4$ directions), $\tilde{B}$ is a (four-dimensional) two-form, $A^i$ are 3 vectors and $\alpha_A,\beta^A$ are 2 three forms on $B_6$. For the solutions we will be interested in, we have that $\xi^A=0=\tilde{\xi}^A $: these are part of the hyperscalars that are set to zero. We will moreover consider purely electric and purely magnetic configurations with zero axions, $b_i =0$.  The M-theory 4-form flux then reads
\be
F_4 = d \mathcal{C}_3 +d \tilde{B} \wedge (\theta + A_0) + H_2^i \wedge \omega_i 
\ee
and where $\theta$ is defined in eq. \eqref{thetadef}. The $\psi$ component of the form $C_{\mu \nu \rho}$, which becomes the NS NS B- field,
\be \label{expr_B}
B = \tilde{B} 
\ee
only has components in four dimensions. The two form $\tilde{B}$ is dual to the massive vector field (see for instance \cite{Gauntlett:2009zw}). The rest of the components of the M-theory form become the Ramond-Ramond 3 form
\be
\label{eq:dC3}
dC_{\alpha \beta \gamma} = H_4 +dB \wedge (\sigma_1+\sigma_2 + \sigma_3+ A_0 )+ H_2^i \wedge \omega_i \,.
\ee

\subsection{DBI action for fluxed D6 branes}

The Lagrangian for the type IIA theory is
\be
S_{IIA} = \frac{2 \pi}{g_s^2 L_s^8} \int d^{10}x \sqrt{-g} \bigg[e^{-2\phi} \left( R +4 (\nabla \phi)^2 - \frac1{12}H_3^2 \right) - \frac12 F_2 \wedge \star F_2  \nonumber
\ee
\be
-\frac12 \tilde{F}_4 \wedge \star \tilde{F}_4 -\frac12 B \wedge F_4 \wedge F_4  \bigg]\,,
\ee
with $F_2 = dC_1$, $\tilde{F}_4 = dC_3 -C_1 \wedge H_3$, $H_3 = dB$ and $F_4 = dC_3$. We follow the conventions of \cite{Myers:1999ps}, which were also adopted in \cite{Asplund:2015yda}.  We will now consider now a D6 brane that wraps the six internal dimensions\footnote{Let us also mention that the bare masses for D-branes  in in $AdS_4 \times CP^3$ compactifications were computed in \cite{Gutierrez:2010bb}.}. We split the DBI action 
into in two parts:
\be
S_{Dp} = S_m+S_{WZ} \,,
\ee
with
\be
\label{eq:WV}
S_{m} = - \frac{2 \pi}{ g_s L_s^{p+1}} \int d^{p+1} e^{-\Phi} \sqrt{-det(g+L_s^2 \mathcal{F})} \,, \qquad S_{WZ} = \frac{2\pi}{g_s L_s^{p+1}} \int \sum_n C_n \wedge e^{L_s^2 \mathcal{F}}
\ee
and
\be
\mathcal{F} =  \frac{1}{L_s^2} B_{\alpha \beta} + \frac{1}{2\pi} dA^{(1)}
\ee
where $A^{(1)}$ is the U(1) gauge field living on the worldvolume, $\Phi$ is the string theory dilaton. As for the embedding ansatz, we have coordinates $\tau, \theta_1,\phi_1,\theta_2, \phi_2, \theta_3,\phi_3$ where $\tau$ goes from $- \infty$ to $+ \infty$ and $ 0 \leq \theta_i < \pi $, $ 0 \leq \phi_i < 2 \pi$.   The D6 brane wraps the entire 6-manifold obtained upon reduction from 11d to IIA, once around each $S^2$, and looks like a particle in 4d.

We take the worldvolume flux to be\footnote{This flux should be quantized in units of $\alpha'$ but we can regard it as a continuous parameter: we consider fluxes of order one, therefore we have a large number of dissolved D- branes (though compatible with the probe approximation).}:
\be \label{worldvol}
L_s^2 \mathcal{F} = \mathbf{f} ( \omega_1 +  \omega_2+  \omega_3 ) \,.
\ee
Looking at the NS NS B-field in \eqref{expr_B}, we notice that for our backgrounds the two form $B$ is
 \be 
dB \propto \star_4 \mathcal{B} \,,
\ee
and the massive vector $\mathcal{B}$ has only component along the time direction. Therefore the $\star_4 \mathcal{B}$ is aligned in $r,\theta,\phi$ directions: this part does not contribute to the action because it is zero on the brane. In addition, the term in $b_i$ is zero due to the absence of axions (also see covariant derivative, as defined in \eqref{covdiv}).
\noindent Summing up the contributions we get
\begin{eqnarray}
S_{m} & = & -2 \pi \int d^{7}x e^{-\Phi} \sqrt{-det \left( g  + \mathbf{f} J \right)}  \nonumber \\
& = & -2 \pi  \int d^{7}x \frac{e^{-\Phi}}{8^3}  \sqrt{ \tau_1^2 \tau_2^2 \tau_3^2 \frac{\left( \tau_1^2 +  \mathbf{f}^2 \right)}{\tau_1^2}  \frac{\left( \tau_2^2+  \mathbf{f}^2 \right)}{\tau_2^2} \frac{\left( \tau_3^2+  \mathbf{f}^2 \right)}{\tau_3^2}   } \sigma \sqrt{g_{tt}} \nonumber \\
& = & -2 \pi^4 \int dt \frac{ \pi^3}{8 \sqrt{\tau_1 \tau_2 \tau_3}} \sqrt{ \left(\tau_1^2+  \mathbf{f}^2 \right)\left( \tau_2^2 +  \mathbf{f}^2 \right)\left( \tau_3^2+  \mathbf{f}^2 \right)}  \sqrt{g_{tt}} \,.
\end{eqnarray}
In passing, we notice the cancellation between the dilaton and the prefactor in the rescaled 10d $g_{tt}$ component. In going from the second to the third line we have integrated over the angular coordinates $\theta_i, \phi_i$: each 2-sphere gives a factor $4\pi$. We can rewrite the formula we have found as
\be
S_m = -2\pi^4  \int dt  \frac{\pi^3}{2^3 \sqrt{\tau_1 \tau_2 \tau_3} } \sqrt{g_{tt}} | (\tau_1+i \mathbf{f}) (\tau_2+i \mathbf{f})(\tau_3+i \mathbf{f})| \,,
\ee
and we notice, as in  \cite{Asplund:2015yda}, that $S_m$ becomes then proportional to the absolute value of the so-called  “complexified K\"ahler volume", obtained by modifying $ J \rightarrow J_{C} \equiv J + i L_s \mathcal{F} $. Towards the end of this section, we will see that this part can be conveniently repackaged in terms of four-dimensional $\mathcal{N}=2$ supergravity quantities.

\vspace{5mm}

Now we have to unpack the Wess Zumino (WZ) piece of the DBI. We denote with $C_i$ the RR forms. We have
\begin{equation*}
S_{WZ} = \frac{2 \pi}{g_s L_s^7 } \left[\int \, C_7 +  L_s^2 \int \, \, C_5 \wedge \mathcal{F} + \frac{L_s^4}2 \int  \, C_3 \wedge  \mathcal{F} \wedge \mathcal{F} + \frac{L_s^6}{6} \int  \, C_1 \wedge \mathcal{F} \wedge  \mathcal{F} \wedge \mathcal{F}  \right]
\end{equation*}
\be
\equiv  S_{WZ,D6} + S_{WZ,D4}+ S_{WZ, D2}+ S_{WZ,D0}
\ee
First of all, the term
\be
S_{WZ, D0} = \frac{2\pi}{g_s L_s } \frac16 \int \, C_1 \wedge \mathcal{F} \wedge  \mathcal{F} \wedge \mathcal{F}  \ee
picks up the 0-th component of the RR field $ C_1$, which upon reduction from 11d is \eqref{Aa}. Given the form of the worldvolume flux \eqref{worldvol}, this term gives rise to this:
\be
S_{WZ, D0}= \frac{2\pi}{g_s L_s^7 } \frac{(4\pi)^3}{8^3} \int dt A^0_t \, \mathbf{f}^3  =  \frac{\pi^4}{ 4 g_s L_s^7 } \int dt A^0_t \, \mathbf{f}^3 \,.
\ee
Secondly, the term
\be
S_{WZ, D2}= \frac{2\pi}{g_s L_s^3 } \frac12 \int ( C_3 \wedge \mathcal{F} \wedge \mathcal{F} )
\ee
where we need to have the zeroth ($t$-) component of $C_3$. It gives
\be
S_{WZ, D2}= -\frac{2\pi}{g_s L_s^7 } \frac{(4\pi)^3 \mathbf{f}^2 }{2 \times 8^3}  \int dt \, 2 \times ( A_t^1 +A_t^2+A_{t}^3)  =  - \frac{\pi^4 \, \mathbf{f}^2}{ 4 g_s L_s^7 } \int dt \, ( A_t^1 +A_t^2+A_{t}^3)\,,
\ee
where we took into account the minus sign coming from \eqref{CC3}.
We then have the piece
\be
S_{WZ, D4}=  \frac{2\pi}{g_s L_s^5 } \int  ( C_5 \wedge \mathcal{F} )\,,
\ee
where $C_5$ is\footnote{The RR dual forms $C_5$ and $C_7$ are defined such that the Bianchi identities for these coincide with the Maxwell's equations for the dual fields $C_3$ and $C_1$. Indeed we define
 \be
\tilde{F}_6   =  dC_5 + C_3 \wedge H_3\,, \qquad 
 \tilde{F}_8  =  dC_7 + C_5 \wedge H_3\,.
 \ee
Demanding that the Bianchi identities $d\tilde{F}_6  =   dC_3 \wedge H_3$ and
$d \tilde{F}_8  =  dC_5 \wedge H_3$ coincide with the  Maxwell's equation of motion for $C_3$ and $C_1$  we get
 \be
 \tilde{F}_6 =- \star_{10}  \tilde{F}_4 \,, \qquad  \star_{10} F_2 = F_8 \,,
 \ee giving in total
 \begin{eqnarray}
 dC_5 &=  & - \star_{10} \, (dC_3 - C_1 \wedge H_3) - C_3 \wedge H_3\,, \\
 dC_7 & =  &  \star_{10} \, dC_1 - C_5 \wedge H_3\,.
 \end{eqnarray}
 Notice that this is consistent with the remarks on the dual pairs after formula (11) in \cite{Myers:1999ps}. }
\be \label{eq:dC5}
dC_5 =  - \star_{10} (dC_3 -   C_1 \wedge H_3) - C_3 \wedge H_3\,.
\ee
This term gives
\be
S_{WZ, D2}=  \frac{2\pi}{g_s L_s^7 } \frac{(4 \pi)^3}{8^3} \int dt \,  \mathbf{f}  \left( \sum_{i=1}^3  B_{i,t} \right) \,,
\ee
where $B_I$ are the magnetic gauge potentials, defined as
\be \label{magn_gauge}
d B_I = G_I \qquad G_I = 2 \frac{\partial \mathcal{L}}{ \partial F^{\mu \nu, I}} 
\ee
Finally,
\be
S_{WZ,D6} = \frac{2\pi}{g_s L_s^7 }  \int  C_7 \,, 
\ee
with
\be
dC_7 =  \star_{10} \, dC_1 - C_5 \wedge H_3
\ee
gives
\be
S_{WZ, D6} = \frac{2\pi}{g_s L_s^7 } \frac{(4 \pi^3)}{8^3} \int dt B_{0,t}\,.
\ee
From these computation, we can seen now that the D-brane charges are
 \be \label{gam}
\gamma  = \left(q_{D6}, q_{D4}, q_{D4}, q_{D4}, q_{D0}, q_{D2}, q_{D2}, q_{D2} \right) = \left( 1, \mathbf{f}, \mathbf{f}, \mathbf{f},\mathbf{f}^3 , \mathbf{f}^2, \mathbf{f}^2, \mathbf{f}^2 \right)\,,
\ee
where the D6 brane corresponds to the supergravity charge $\mathbf{P}^0$, the induced charges of the D4 branes wrapped on different 4-cycles of $B_6$ correspond to the supergravity probes charges $\mathbf{P}^1, \mathbf{P}^2, \mathbf{P}^3$, D2 branes wrapped on different 2-cycles of $B_6$ are $\mathbf{Q}_1, \mathbf{Q}_2, \mathbf{Q}_3$ and the D0 brane charge in the 4d supergravity\footnote{The minus sign in $q_{D0} = -\mathbf{Q}_0$ is due to a trivial difference in conventions and was already noticed in \cite{Asplund:2015yda}, Sec. 4.3.} conventions corresponds to $-\mathbf{Q}_0$. We have verified that this charge assignment can also be obtained from the lower dimensional supergravity theory by requiring that the discriminant $\mathcal{D}(\gamma)$ (also called "quartic invariant" $I_4$) is invariant under the shift of charges described in Sect. 4.1 of \cite{Anninos:2011vn}.

At the end of the day the WZ part of the action takes the form
\begin{equation*}
S_{WZ} = S_{WZ,D0}+S_{WZ,D2}+S_{WZ,D4}+S_{WZ,D6}  =
\end{equation*}
\be
 =  \frac{\pi^4}{ 4 g_s L_s^7 } \int dt  \left( B_{0,t} \times 1 + \left( \sum_{i=1}^3 B_{i,t} \right) \times \mathbf{f} \textcolor{blue}{-} \left( \sum_{i=1}^3 A_{t}^i \right) \times \mathbf{f}^2 + A_{t}^0 \times \mathbf{f}^3\right)\,.
 \ee
 This can be rewritten in a more compact form by defining the four-dimensional symplectic vector of electromagnetic charges
 \be \label{chargeprobe_4d}
\Gamma = (\mathbf{P}^0, \mathbf{P}^1,\mathbf{P}^2,\mathbf{P}^3, \mathbf{Q}_0,\mathbf{Q}_1,\mathbf{Q}_2,\mathbf{Q}_3) = \left( 1, \mathbf{f}, \mathbf{f}, \mathbf{f},- \mathbf{f}^3 , \mathbf{f}^2, \mathbf{f}^2, \mathbf{f}^2 \right)\,.
 \ee
 In this way, the full $S_{WZ}$ reads
 \be
S_{WZ}  = \frac{\pi^4}{ 4 g_s L_s^7 } \int  (- \mathbf{Q}_I A^I +  \mathbf{P}^I B_I) = - \frac{\pi^4}{ 4 g_s L_s^7 } \int  \langle \mathbb{A} ,  \Gamma  \rangle \,,
\ee
where $ \mathbb{A} = (A^I, B_I) $ is the symplectic vector of the electric and dual magnetic potentials, $B_I$ being the dual potentials to the gauge fields $A^I$ defined as in \eqref{magn_gauge} and the probe charge vector $\gamma$ is defined in \eqref{gam}. The symplectic product between two vectors  $m,n$, such that $ m= (m^I,m_I) $, $n= (n^I,n_I)$, is defined as
\be
 \langle m, n \rangle =  m^I n_I  - n^I m_I \,.
\ee
We can now repackage the first term, $S_m$ in the DBI action, which reads 
\begin{equation*}
\frac{1}{2 \sqrt2 \sqrt{\tau_1 \tau_2 \tau_3}} | (\tau_1+i \mathbf{f}) (\tau_2+i \mathbf{f})(\tau_3+i \mathbf{f}) | = 
\end{equation*}
\be
=\frac{1}{2 \sqrt2 \sqrt{\tau_1 \tau_2 \tau_3}} \left(\mathbf{f}^6 +\mathbf{f}^4 (\tau_1^2 + \tau_2^2+\tau_3^2)+ \mathbf{f}^2 (\tau_1^2 \tau_2^2+ \tau_3^2 \tau_1^2+ \tau_2^2 \tau_3^2) +\tau_1^2 \tau_2^2 \tau_3^2 \right)^{1/2} \label{total_ccc}
\ee
in terms $ \mathcal{Z}  \equiv \langle \mathcal{V} , \Gamma \rangle $, the central charge of the $\mathcal{N} =2$ theory, defined by the symplectic product of the vector of the electromagnetic charges $\Gamma$ and the covariantly holomorphic sections $\mathcal{V}$ defined in \eqref{cov_holom_sect}. Using \eqref{chargeprobe_4d}, this reads
\be
|\mathcal{Z}(\tau^i, \Gamma)|^2 = \frac{1}{8 \tau_1 \tau_2 \tau_3} \left(\mathbf{f}^2 \tau_2^2 \tau_1^2+\mathbf{f}^2 \tau_2^2 \tau_3^2+\mathbf{f}^4 \tau_2^2+\mathbf{f}^2 \tau_1^2 \tau_3^2+\mathbf{f}^4 \tau_1^2+\mathbf{f}^4 \tau_3^2+\mathbf{f}^6+\tau_2^2 \tau_1^2 \tau_3^2 \right)\,,
\ee
which matches with \eqref{total_ccc}. In total for $S_m$ we then have
\be
S_m = - \frac{\pi^4}{\sqrt2} \int dt \, \sqrt{g_{tt}} \, | \mathcal{Z}(\tau^i, \Gamma)|\,.
\ee
Apart from overall normalization factors, this expression reproduces formula (4.11) of \cite{Asplund:2015yda}: there it is shown that the gravitational part of the DBI action for a fluxed D6 brane wrapped around $\mathbb{CP}^3$ becomes proportional to the absolute value of the central charge $|\mathcal{Z}|$. Here we have shown that the same is valid for a fluxed D6 brane wrapping the 6d manifold obtained by the reduction on $\psi$ of the compactification on $Q^{111}$. 

Hence in total we arrived at the following probe action:
\be \label{total_D6_probe}
S =  - \frac{\pi^4}{\sqrt2} \int dt \, \sqrt{g_{tt}} \, |\mathcal{Z}(\tau^i, \Gamma)|- \frac{\pi^4}{ 4  } \int  \langle \mathbb{A}, \Gamma \rangle\,.
\ee

In asymptotically flat spacetimes the energy of a BPS state is obtained as its eigenvalue under the action of the central  charge  operator  appearing  in  the $\mathcal{N} =2$ superalgebra. In AdS$_4$ spacetimes this we cannot directly apply this reasoning since there is no such central charge operator in any of the two possible superalgebras allowed in $\mathcal{N}=2$ gauged supergravity \cite{Hristov:2011ye}. As noticed in \cite{Asplund:2015yda}, it is nevertheless remarkable that the form of the probe action obtained from the DBI resembles the one of flat space.

As a final remark, let us mention that the total probe action \eqref{total_D6_probe} we obtained (written in the frame with $stu$ prepotential) reduces\footnote{Apart from some normalization factors in the charges, due to the fact that in \cite{Asplund:2015yda} the chosen model has prepotential  $F\propto \frac{(X^1)^3}{X^0}$, namely sections and gauge fields are identified from the start, while in our model they are kept different and are identified only a posteriori.} to formula (4.11) of  \cite{Asplund:2015yda} for the choice $X^1 =X^2=X^3$, and probe charges $\mathbf{Q}_1=\mathbf{Q}_2=\mathbf{Q}_3$ and $\mathbf{P}^1 = \mathbf{P}^2 = \mathbf{P}^3$. Their setup corresponds moreover to the black hole background charges $Q_0 = P^i=0$, $i=1,2,3$.

\section{\texorpdfstring{$Q^{111}$}{Q111} black hole solutions with Supersymmetry }
\label{susyHPZ}

We give here a brief summary of the details of the supersymmetric solutions of Halmagyi-Petrini-Zaffaroni \cite{Halmagyi:2013sla}\footnote{Subsets of the solution space were found also in \cite{Donos:2008ug,Donos:2012sy}. See \cite{Hong:2019wyi,Kim:2019umc,Kim:2020qec} for other studies regarding supersymmetric AdS$_2$ near horizon geometries in Sasaki--Einstein truncations and black hole microstates.}. The latter paper analytically constructs the near-horizon AdS$_2 \times \Sigma_g $ geometry of supersymmetric black holes in various $SE_7$ truncations, including the models arising from  $Q^{111} $ and $M^{111}$, in the $mSTU$ prepotential. The geometry interpolating between the near horizon geometry and the asymptotically AdS$_4$ space is constructed numerically by solving the ODEs from the Killing spinor equations. 

The conditions to preserve supersymmetry are
\begin{eqnarray}
P^{\Lambda} P_{\Lambda}^3 & = & - \kappa \,, \\
P^{\Lambda} k_{\Lambda}^u &= & 0\,,
\end{eqnarray}
and
\begin{eqnarray}
\mathcal{L}_r^{\Lambda} P_{\Lambda}^3 &= &0 \,, \\
\mathcal{L}_i^{\Lambda} k_{\Lambda}^u &= &0\,,
\end{eqnarray}
where $\mathcal{L}^{\Lambda}$ is
\be
\mathcal{L}^{\Lambda} = \mathcal{L}_r^{\Lambda} + i \mathcal{L}_i^{\Lambda} = - e^{\mathcal{K}/2} X^{\Lambda}\,.
\ee
For concreteness, we will focus on the $M^{111}$ model with Freund-Rubin parameter\footnote{Let us remind the reader that our conventions are such that $e_{0,HPZ} = 6 e_{0,ours}$} $e_{0,HPZ} =6$. Inserting the values of the prepotentials $P_{\Lambda}^3$ and the quaternionic Killing vectors $k_{\Lambda}^u$ 
\be
P_{\Lambda}^{3}= \sqrt2 \{ 4- \frac12 e^{2\phi} e_0, -2e^{2\phi}, -e^{\phi} \} \,,
\ee
\be
k_{\Lambda}^a= -\sqrt2 \{ e_0, 4,2 \}\,.
\ee
The NH geometries of \cite{Halmagyi:2013sla} are of the form
\be
ds^2 =  - R_1^2 r^2 dt^2 + \frac{dr^2}{ r^2 R_1^2} + R_2^2 (d\theta^2 + f(\theta)^2 d\phi^2)\,, \qquad \quad f(\theta) = \left \{ \begin{array}{cc} 
\sin \theta \,\,\, \text{for} \,\,\, \kappa =1 \\
\sinh \theta \,\,\, \text{for} \,\,\, \kappa =-1 \end{array} \right.
\ee
with gauge fields
 \be
 A^{\Lambda}=  \tilde{q}^{\Lambda} r dt + P^{\Lambda} f'(\theta) d \phi\,,
 \ee
 \be
\tilde{q}^{\Lambda} = -\frac{1}{R_2^2} \left(   \im \cN^{\Lambda \Sigma}  \re \cN_{\Sigma \Gamma} P^{\Gamma}+ \im \cN^{\Lambda \Sigma} Q_{\Sigma} \right)\,,
 \ee
 and constant values of the scalars. The conditions for supersymmetry leave a two-dimensional parameter space given by 
\be
b_3  =  b_1\,, \qquad \tau_3 =\tau_1\,, \qquad P^3 = P^1\,, \qquad Q_3 = Q_1\,,
\ee
with
\begin{eqnarray} \label{firstset}
b_1 & = & \epsilon_2 \sqrt{\frac{\tau_1 (6- 2 \tau_1 (\tau_1 + 2\tau_2))}{2 (\tau_1 + 2\tau_2)}} \,,\nonumber \\
b_2 & = & - \frac{(\tau_1 +\tau_2)b_1}{\tau_1} \,, \nonumber \\
e^{2\phi} & = & \sigma^2 =  \frac{4(\tau_1 +2 \tau_2)^2}{2 \tau_1^4 + 8 \tau_1^3 \tau_2 + (18 + 8 \tau_1^2) \tau_2^2} \,, \nonumber \\ 
R_1 & = & \frac{\tau_1 \sqrt{\tau_2}}{4} \,, \nonumber \\
R_2^2 & = & \kappa R_1^2 \frac{(2 \tau_1^4 + 8 \tau_1^3 \tau_2 + (18 +8 \tau_1^2) \tau_2^2)}{\tau_2 (18 \tau_2 -4 \tau_1 (\tau_1+2\tau_2)^2)}\,,
\end{eqnarray}
and charges
\begin{eqnarray} \label{secondset}
P^0 & = & -\frac{1}{4 \sqrt2}\,,  \nonumber \\
P^2 & = & \frac{3}{4\sqrt2} -2 p^1 \,, \nonumber \\
P^1 & = & - \frac{3}{4 \sqrt2} \frac{2\tau_1^4-18 \tau_2 (\tau_1 + \tau_2) + 12 \tau_1^2 \tau_2 (\tau_1 + 2\tau_2) +16 \tau_1 \tau_2^3 }{(\tau_1+2\tau_2)(18 \tau_2 -4\tau_1(\tau_1+2\tau_2)^2)} \,, \nonumber \\
Q_0& = & - \frac{\epsilon_2 \kappa}{16} \sqrt{\frac{\tau_1 (6- 2 \tau_1 (\tau_1 + 2\tau_2))}{2 (\tau_1 + 2\tau_2)}}  \nonumber \\
& & \times \left(  \frac{8 \tau_1^6 - \tau_2(\tau_1 +\tau_2) (108 + 24 \tau_1^2 -48 \tau_1^4) +48 \tau_1^4 \tau_2^2 + 8 \tau_1 \tau_2^3 (6 + 8 \tau_1^2)  }{18\tau_2 -4\tau_1(\tau_1 +2\tau_2)^2} \right)\,,\nonumber \\ 
Q_1 & = &- \frac{\epsilon_2 \kappa}{8} \sqrt{\frac{\tau_1 (6- 2 \tau_1 (\tau_1 + 2\tau_2))}{2 (\tau_1 + 2\tau_2)}} \frac{18 \tau_2 -2 \tau_1 (\tau_1 + 2\tau_2)^2}{18\tau_2 -4\tau_1(\tau_1 +2\tau_2)^2} \,,\nonumber \\
Q_2 & = & - \frac{\epsilon_2 \kappa }{8} \sqrt{\frac{\tau_1 (6- 2 \tau_1 (\tau_1 + 2\tau_2))}{2 (\tau_1 + 2\tau_2)}} \frac{4 \tau_1^4 +\tau_2 (16 \tau_1^2-18)(\tau_2+\tau_1)}{18\tau_2 -4\tau_1(\tau_1 +2\tau_2)^2} \,.
\end{eqnarray}
The supersymmetry conditions allow for the axions $b_i$ to vanish and for purely magnetic solutions. One choice is this
\be
6- 2 \tau_1 (\tau_1 + 2\tau_2) = 0 \qquad \rightarrow \qquad \tau_2 = \frac{3- \tau_1^2}{2 \tau_1}\,.
\ee 
The near horizon solution is then
\be
Q_0= Q_2 = Q_3 =0 \,, \qquad P^0 = -\frac{1}{4 \sqrt2}\,, \qquad P^2 = \frac{3}{4\sqrt2} -2 P^1 \,, \qquad 
\ee
with
\be \label{chargeVSscalar}
P^1 =  -\frac{ \tau_1^2 \left( \tau_1^2-3\right)}{4 \sqrt{2} \left(\tau_1^2+1\right)} \,,
\ee
and\footnote{We thank Hyojoong Kim for pointing out a typo (which did not propagate in the subsequent steps) in the value of $R_2$ in the first version of the manuscript.}
\be
R_1^2 =\frac{\tau_1}{32}  \left( 3 - \tau_1^2\right) \,, \qquad R_2^2 = -\frac{\kappa \tau_1 \left(\tau_1^4-2 \tau_1^2+9\right)}{32 \left(\tau_1^2+1\right)} \,.
\ee
Solutions with positive values of $R_1$ and $R_2$ and scalars inside the K\"ahler cone exist, however they need $\kappa =-1$ (for instance, $\kappa =-1, \tau_1 = 1.14173$ gives $R_1^2 \approx 0.06 $ and $R_2^2 \approx 0.125 $). As a consistency check, we have verified that these configurations are solutions to our equations of motion\footnote{Our conventions differ from those of \cite{Halmagyi:2013sla} by a factor in the definition of the charges, i.e. $P^{\Lambda}_{HPZ} = \frac{1}{\sqrt2} P^{\Lambda}_{ours} $ }. In the main text, we show that we have constructed full-flow solutions with sets of charges  satisfying \eqref{firstset}-\eqref{secondset} and close to extremality from or system of second order equations of motion.

  \section{Exact solutions from the \texorpdfstring{$S^7$}{S7} truncation}
      \label{sec:S7comparison}
      
Analytic AdS$_4$ black hole solutions exist in $\mathcal{N}=2$, $U(1)$-gauged supergravity in the absence of hypermultiplets (with Fayet-Iliopoulos gauging). This theory comes from the reduction of 11 dimensional supergravity on $S^7$ \cite{deWit:1986oxb} and is characterized by prepotential $F = -2i \sqrt{X^0 X^1 X^2 X^3}$. The black holes are found as solutions to the equations of motion of the Lagrangian
\begin{eqnarray}
\mathcal{L} & = & \frac{R}{2} +  g_{i\bar{\jmath}}Dt^i\wedge\ast D\bar{t}^{\bar{\jmath}}
       +  \tfrac{1}{4}\im\mathcal{N}_{\Lambda \Sigma}F^{\Lambda}\wedge\ast F^{\Sigma}
      + \tfrac{1}{4}\re\mathcal{N}_{\Lambda \Sigma}F^{\Lambda}\wedge F^{\Sigma} + V \,,
\end{eqnarray}
with the same $g_{i\bar{\jmath}}$ and $\mathcal{N}_{\Lambda \Sigma}$ as in \eqref{GIJ} and \eqref{matrixN}. Assuming $\tau^i$ real and positive, the potential reads
\be
V = - 2 g^2 \left(\frac{1}{\tau_1} + \tau_1 + \frac{1}{\tau_2} + \tau_2+\frac{1}{\tau_3} + \tau_3  \right) \,.
\ee
The purely electric solutions that we are interested in are of the form \cite{Duff:1999gh}
\be \label{ansatz_static}
ds^2 = - U^2(r) dt^2 + \frac{dr^2}{ U^2(r)} + h^2(r) (d\theta^2 + \sin^2 \theta d\phi^2 ) \,,
\ee
with warp factors
\be
h^2(r) = r^2 \sqrt{H_0 H_1 H_2 H_3} \,, \qquad H_{\Lambda} = 1 + \frac{b_{\Lambda}}{r} \,,
\ee
\be
U^2(r) = \frac{1}{\sqrt{H_0 H_1 H_2 H_3}} \left( 1 -\frac{\mu}{r} + 2g^2 r^2 H_0 H_1 H_2 H_3 \right) \,,
\ee
and purely electric gauge fields
\be
 F_{rt}^{\Lambda} = \frac{1}{2 \sqrt2 h^2} \im\mathcal{N}^{\Lambda \Sigma} Q_{\Sigma} \,,
\ee
scalars
\be
\tau_1 = \sqrt{\frac{H_2 H_3}{H_0 H_1}} \,, \qquad \tau_2 = \sqrt{\frac{H_1 H_3}{H_2 H_0}} \,, \qquad \tau_3 = \sqrt{\frac{H_2 H_1}{H_0 H_3}} \,,
\ee
with parameters
\be \label{final_parameters}
b_{\Lambda} = \mu \sinh^2(q_{\Lambda}) \,,  \qquad Q_{\Lambda} = \mu \sinh(q_{\Lambda}) \cosh(q_{\Lambda})\,.
\ee
These solutions can be uplifted to 11-dimensional supergravity on the seven-sphere and they can be interpreted as spinning M2 branes \cite{Cvetic:1999xp}. Notice that the black hole horizon is located at the largest real zero of the function $U^2(r)$ and the singularity is at the zero of $H_0 H_1H_2H_3$ which is usually located at a finite nonzero radial coordinate. When the gauging is turned off $g=0$ these solutions become non extremal black holes with Minkowski asymototics, solutions of theories of ungauged supergravity. Upon further setting $\mu =0$ and $q_{\Lambda} = b_{\Lambda}$, we retrieve the 1/2 BPS black hole solutions of \cite{Behrndt:1997ny}. 

\begin{figure}[H]
\begin{center}
    \includegraphics[width=90mm]{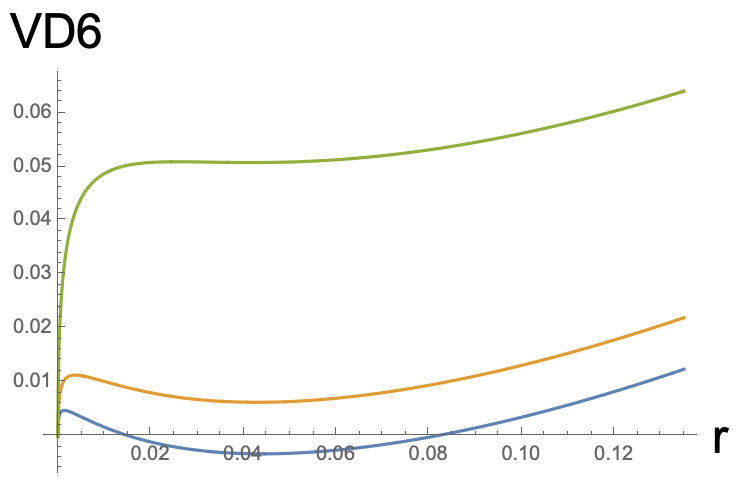}
    \end{center}
    \caption{Example of fluxed D6 brane potential for the spherical AdS$_4$ solution from the $S^7$ truncation. The plot is for the parameters $l_{AdS} = 1/2$,  $T = 0.05 (\text{blue}), 0.11 (\text{orange}), 0.38 (\text{green})$, with probe charge $\kappa =0.44$. For each of the solutions we have shifted the radial coordinate in such a way that the horizon is at $r=0$. \label{figure_Fayet}}
  \end{figure}

The thermodynamics of these solutions was studied in \cite{Cvetic:1999ne} and, in a different symplectic frame, in \cite{Anninos:2013mfa}. In the latter paper, a detailed analysis of stability of probe D-branes on the black hole background, which provides a simple adaptation of the D0-D4 system to asymptotically AdS spacetimes, was performed. The probe action \cite{Billo:1999ip}, is again composed of two parts
\be
V_{D6} = V_g + V_e \,,
\ee
with
\be
V_g = \sqrt{ g_{tt}} \,  \, |\mathcal{Z} (\Gamma, \tau^i)| \,, \qquad V_e = Q_{\Lambda} A^{\Lambda} - P^{\Lambda} B_{\Lambda} \,,
\ee
where we have used the same notation as in \eqref{probe_final}. This form for the probe action, originally assumed in \cite{Anninos:2013mfa}, was later shown to be  can be derived from the DBI action of a fluxed D6 brane in \cite{Asplund:2015yda}. One of the main messages on \cite{Anninos:2013mfa} and the previous related studies in asymptotically flat spacetimes  \cite{Anninos:2011vn} is that the black hole bound states initially found by \cite{Denef:2000nb,Bates:2003vx} in the form of BPS configurations persist at finite temperature, even in asymptotically AdS spacetimes. Indeed there are regions of parameter space where configurations of stable and metastable probes exist in the background of the black holes \eqref{ansatz_static} -\eqref{final_parameters}, the precise region was spelled out in detail in the grand canonical ensemble. It is relatively easy to translate their results in the canonical ensemble (fixed charge), and one can indeed provide examples of stable and metastable probes such as those in Figure \ref{figure_Fayet}. 

For our purposes, the study in the canonical ensemble that we have illustrated in this section was instrumental in providing an adequate guess on the region of parameters in which one could reasonably expect stable probes. Thanks to this intuition we could find the stable probes of Section \ref{sec_D6probes}, see Fig. \ref{Fig_profile_potential}. This notwithstanding, finding stable probes in our model was a nontrivial task, due to the fact that we had to take into account also the presence of hypermultiplets, which give mass to one of the vectors, and this made the numerical work challenging.

As a last remark, we point out that there exist also purely magnetic solutions with a regular supersymmetric limit \cite{Cacciatori:2009iz,DallAgata:2010ejj,Hristov:2010ri} and without axions. The solution can have spherical, planar and hyperbolic horizons, and reads
\be
U^2(r) = \frac{1}{\sqrt{H_0 H_1 H_2 H_3}} \left( 2g^2 r^2 + c_1 -\frac{\mu}{r}+ \frac{c_2}{r^2} \right)  \,,
 \qquad h^2(r) = r^2 \sqrt{H_0 H_1 H_2 H_3}  \,,
\ee
with
\be
\qquad H_{\Lambda} = 1 + \frac{b_{\Lambda}}{r}  \,, \qquad  F_{\theta \phi}^{\Lambda} = \frac{1}{2 \sqrt2} P^{\Lambda} f'(\theta) \,,
\ee
scalars
\be
\tau_1 = \sqrt{\frac{H_0 H_1}{H_2 H_3}} \,, \qquad \tau_2 = \sqrt{\frac{H_2 H_0}{H_1 H_3}} \,, \qquad \tau_3 = \sqrt{\frac{H_0 H_3}{H_2 H_1}} \,,
\ee
with parameters
\be
\sum_{\Lambda=0}^3 b_{\Lambda} =0 \,, \qquad c_1 = \kappa+ 2 b_0 g^2 (b_1+b_2+b_3)+2 b_1 g^2 (b_2+b_3)+2 b_2 b_3 g^2 \,.
\ee
Eliminating one of the scalar parameters, for example $b_3$, we obtain
\begin{eqnarray}
(P^0)^2  &= & - b_0^2 \left(2 b_1^2 g^2+2 b_1 b_2 g^2+2 b_2^2 g^2+1\right)-2 b_0^3 g^2 (b_1+b_2)+ b_0 \mu+c_2 \,, \nonumber \\
(P^1)^2 & = & -b_1^2 \left(2 b_0^2 g^2+2 b_0 b_2 g^2+2 b_2^2 g^2+1\right)-2 b_1^3 g^2 (b_0+b_2)+b_1 \mu+ c_2 \,, \nonumber \\
(P^2)^2 & = & - b_2^2 \left(2 b_0^2 g^2+2 b_0 b_1 g^2+2 b_1^2 g^2+1\right)-2 b_2^3 g^2 (b_0+ b_1)+ b_2 \mu+ c_2 \,, \nonumber \\
(P^3)^2 & = & 2 g^2 (b_0+b_1+b_2)^2 (b_0 (b_1+b_2)+b_1 b_2)   -b_0^2-2 b_0 (b_1+b_2)-b_1^2 + \nonumber \\
&- &2 b_1 b_2-b_2^2+ c_2 - \mu (b_0+b_1+b_2) \,, \nonumber \\
\end{eqnarray}
with supersymmetric limit obtained when
\be
c_2 = \frac{c_1^2}{8 g^2}\,.
\ee
Scanning the parameter space to the best of our possibilities, we could not find stable or metastable probes with these backgrounds. This result goes in the same direction as Sec. \ref{susy-magn}, denoting the absence of stable probes on these sort of supersymmetric magnetic backgrounds.

  \end{appendices}

  \bibliographystyle{utphys}
  \bibliography{references}

\end{document}